\documentclass[final]{siamltex}
\usepackage{graphics}
\usepackage{amssymb}
\usepackage{amsmath}
\usepackage[numbers,sort&compress]{natbib}

\renewcommand{\exp}[1]{e^{#1}}
\renewcommand{\vec}[1]{\mathbf{#1}}

\newcommand{\order}{\mathcal{O}}
\newcommand{\cDp}{\mathcal{D}^+}
\newcommand{\cDm}{\mathcal{D}^-}
\newcommand{\cDpm}{\mathcal{D}^\pm}
\newcommand{\cPp}{\mathcal{P}^+}
\newcommand{\cPm}{\mathcal{P}^-}
\newcommand{\cPpm}{\mathcal{P}^\pm}
\newcommand{\cQp}{\mathcal{Q}^+}
\newcommand{\cQm}{\mathcal{Q}^-}
\newcommand{\cQpm}{\mathcal{Q}^\pm}
\newcommand{\brho}{{\bar \rho}}
\newcommand{\trho}{{\tilde \rho}}
\newcommand{\tmu}{{\tilde \mu}}
\newcommand{\Vee}{{\cal{V}}}
\newcommand{\Wee}{{\cal{W}}}
\newcommand{\Op}{{\Omega_{\brho}}}
\newcommand{\Om}{{\Omega}}
\newcommand{\F}{{G}}

\newcommand{\bsrho}{{{\brho_*}}}

\DeclareFontFamily{OT1}{pzc}{}
\DeclareFontShape{OT1}{pzc}{m}{it}{<-> s * [1.10] pzcmi7t}{}
\DeclareMathAlphabet{\mathpzc}{OT1}{pzc}{m}{it}

\newcommand{\q}{\mathpzc{q}}

\title{A Primer of Swarm Equilibria}
\author{Andrew Bernoff \footnotemark[2]\ \footnotemark[4] \and Chad M. Topaz \footnotemark[3]\ \footnotemark[5]}

\begin{document}

\maketitle
\renewcommand{\thefootnote}{\fnsymbol{footnote}}
\footnotetext[2]{Department of Mathematics, Harvey Mudd College, Claremont, California 91711 USA}
\footnotetext[3]{Department of Mathematics, Statistics, and Computer Science, Macalester College, St. Paul, Minnesota 55105 USA}
\footnotetext[4]{ajb@hmc.edu}
\footnotetext[5]{ctopaz@macalester.edu}
\renewcommand{\thefootnote}{\arabic{footnote}}

\begin{abstract}
We study equilibrium configurations of swarming biological organisms subject to exogenous and pairwise endogenous forces. Beginning with a discrete dynamical model, we derive a variational description of the corresponding continuum population density. Equilibrium solutions are extrema of an energy functional, and satisfy a Fredholm integral equation. We find conditions for the extrema to be local minimizers, global minimizers, and minimizers with respect to infinitesimal Lagrangian displacements of mass. In one spatial dimension, for a variety of exogenous forces, endogenous forces, and domain configurations, we find exact analytical expressions for the equilibria. These agree closely with numerical simulations of the underlying discrete model.The exact solutions provide a sampling of the wide variety of equilibrium configurations possible within our general swarm modeling framework. The equilibria typically are compactly supported and may contain $\delta$-concentrations or jump discontinuities at the edge of the support. We apply our methods to a model of locust swarms, which are observed in nature to consist of a concentrated population on the ground separated from an airborne group. Our model can reproduce this configuration; quasi-two-dimensionality of the model plays a critical role.
\end{abstract}

\begin{keywords}
swarm, equilibrium, aggregation, integrodifferential equation, variational model, energy, minimizer, locust
\end{keywords}

\pagestyle{myheadings}
\thispagestyle{plain}
\markboth{A.J. BERNOFF AND C.M. TOPAZ}{A PRIMER OF SWARM EQUILIBRIA}

\section{Introduction}
\label{sec:intro}

Biological aggregations such as fish schools, bird flocks, bacterial colonies, and insect swarms \cite{ParHam1997,BenCohLev2000,OkuLev2001} have characteristic morphologies governed by the group members' interactions with each other and with their environment. The \emph{endogenous} interactions, \emph{i.e.}, those between individuals, often involve organisms reacting to each other in an attractive or repulsive manner \cite{Bre1954,MogEde1999} when they sense each other either directly by sound, sight, smell or touch, or indirectly via chemicals, vibrations, or other signals. A typical modeling strategy is to treat each individual as a moving particle whose velocity is influenced by social (interparticle) attractive and repulsive forces \cite{MogEdeBen2003,LukLiEde2010}. In contrast, the \emph{exogenous} forces describe an individual's reaction to the environment, for instance a response to gravity, wind, a chemical source, a light source, a food source, or a predator. The superposition of endogenous and exogenous forces can lead to characteristic swarm shapes; these equilibrium solutions are the subject of our present study.

\begin{figure}
\resizebox{\textwidth}{!}{\includegraphics{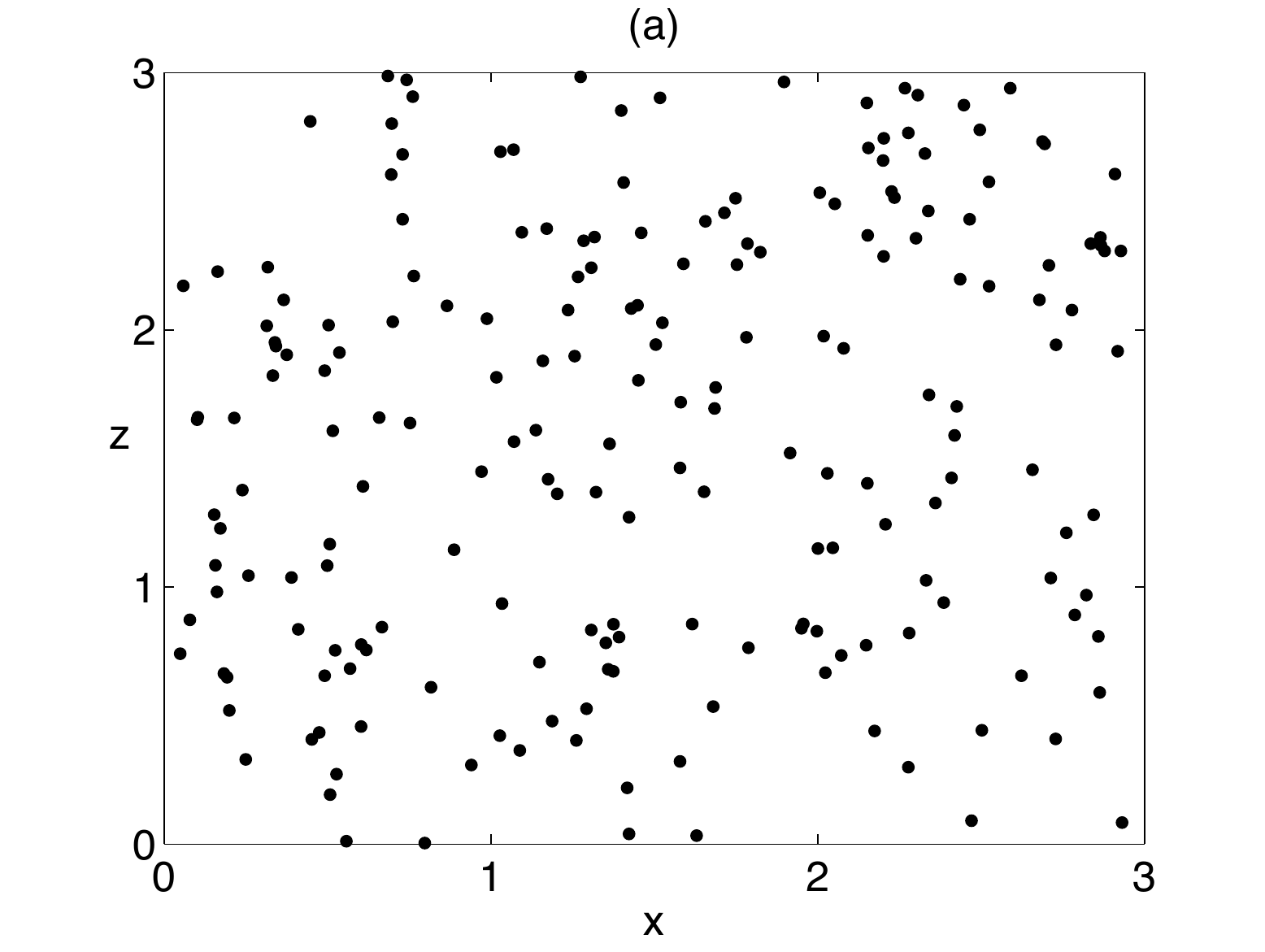} \includegraphics{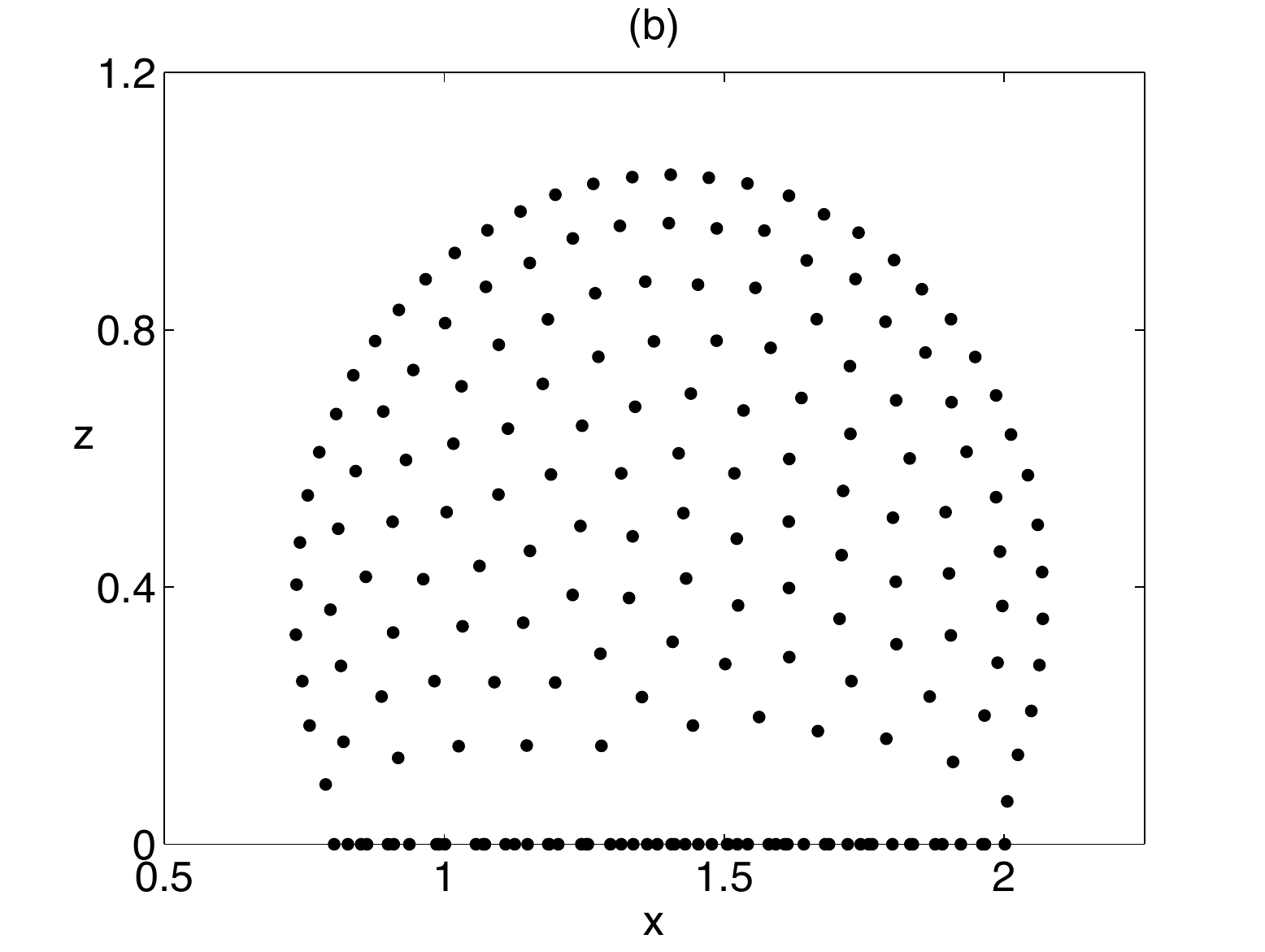}}
\resizebox{\textwidth}{!}{\includegraphics{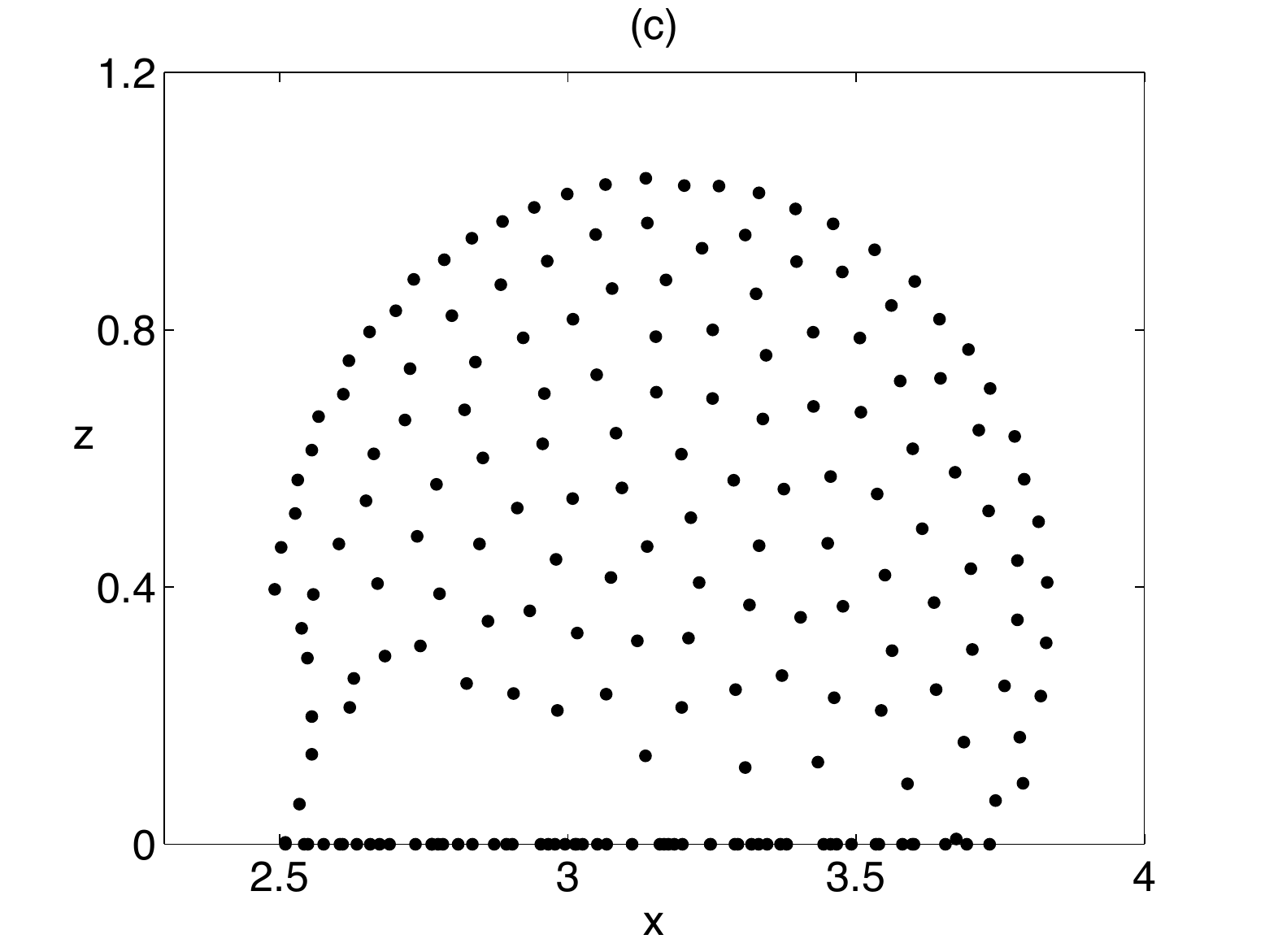} \includegraphics{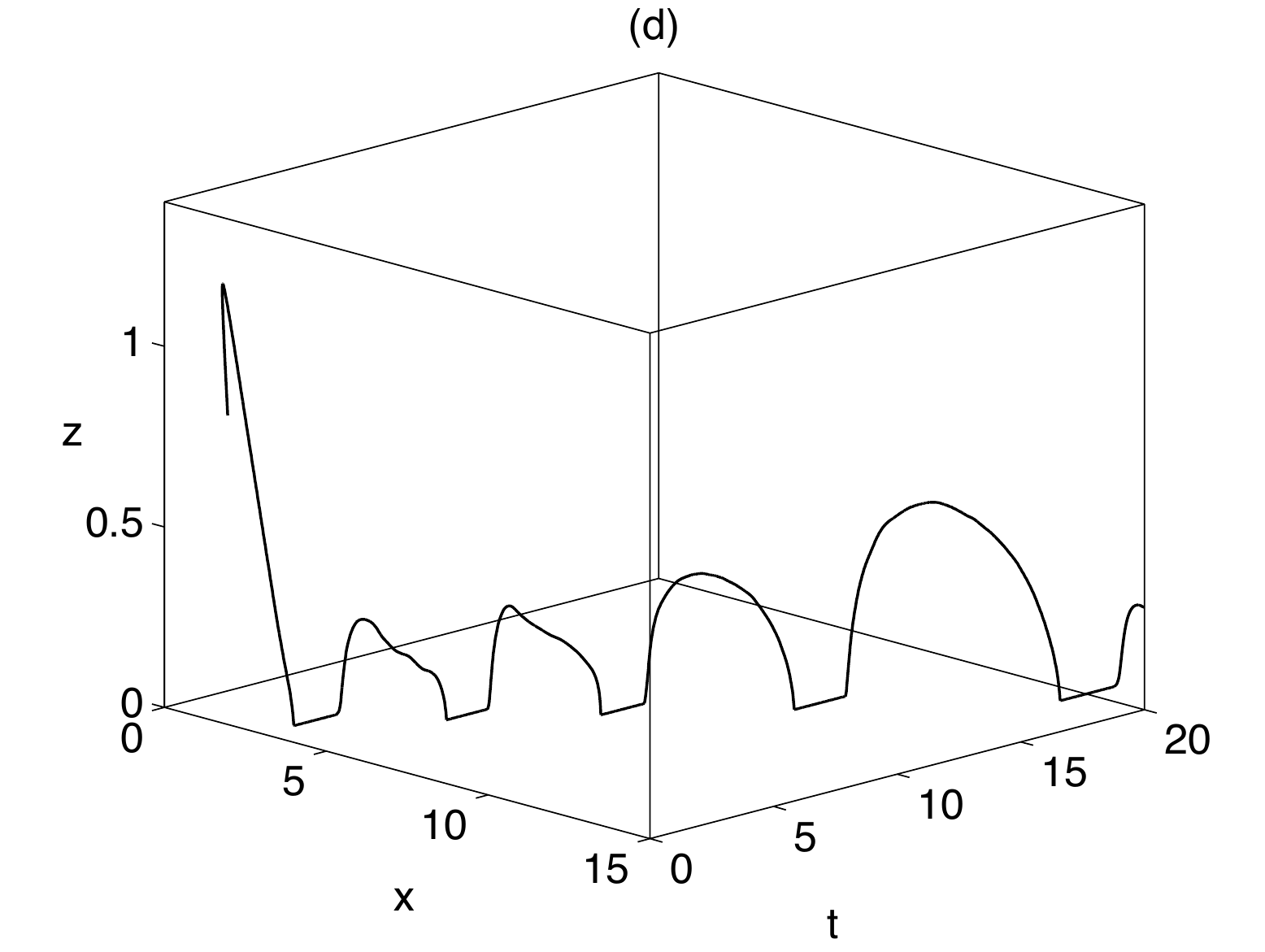}}
\caption{Numerical simulation of the two-dimensional locust swarm model of~\cite{TopBerLog2008}, given by~(\ref{eq:locusts}).~(a) The simulation begins with a randomly distributed initial state.~(b) With gravity but no wind ($U=0$), the swarm's equilibrium is a bubble-like shape on the ground, consisting of a dense, grounded group of locusts and an airborne group. The two are separated by a gap that is void of insects. (c) For the full simulation with wind, by the time $t=2$, the swarm again coheres into a bubble and travels to the right with a rolling motion. Individuals in the back of the swarm take off to join the flying group and individuals reaching the front of the flying swarm land on the ground, where they remain motionless until taking off again.~(d) The trajectory of one individual locust from $t = 0$ to $t=20$ demonstrates the periodic landing and takeoff. The parameters in (\ref{eq:locusts}) are $N=200$, $\F=0.5$, $L=10$, $g=1$ and $U=1$. \label{fig:locust}}
\end{figure}

More specifically, our motivation is rooted in our previous modeling study of the swarming desert locust \emph{Schistocerca gregaria} \cite{TopBerLog2008}. In some parameter regimes of our model (presented momentarily), locusts self-organize into swarms with a peculiar morphology, namely a bubble-like shape containing a dense group of locusts on the ground and a flying group of locusts overhead; see Figure \ref{fig:locust}(bc). The two are separated by an unoccupied gap. With wind, the swarm migrates with a rolling motion. Locusts at the front of the swarm fly downwards and land on the ground. Locusts on the ground, when overtaken by the flying swarm, take off and rejoin the flying group; see Figure \ref{fig:locust}(cd). The presence of an unoccupied gap and the rolling motion are found in real locust swarms ~\cite{Alb1967,Uva1977}. As we will show throughout this paper, features of swarms such as dense concentrations and disconnected components (that is, the presence of gaps) arise as properties of equilibria in a general model of swarming.

The model of \cite{TopBerLog2008} is
\begin{subequations}
\label{eq:locusts}
\begin{gather}
\dot{\vec{x}}_i = \left( \sum_{j=1}^{N} \q(|\vec{r}_{ij}|) \frac{\vec{r}_{ij}}{|\vec{r}_{ij}|} \right)  -g \hat{\vec{e}}_z + U \hat{\vec{e}}_x ,\quad i=1 \ldots N, \\
\q(r) = \F \exp{-r/L} - \exp{-r}, \quad \vec{r}_{ij} = \vec{x}_j - \vec{x}_i,
\end{gather}
\end{subequations}
which describes $N$ interacting locusts with positions $\vec{x}_i$. The direction of locust swarm migration is strongly correlated with the direction of the wind \cite{Uva1977,Rai1989} and has little macroscopic motion in the transverse direction, so the model is two-dimensional, \emph{i.e.}, $\vec{x}_i = (x_i,z_i)$ where the $x$ coordinate is aligned with the main current of the wind and $z$ is a vertical coordinate. As the velocity of each insect is simply a function of position, the model neglects inertial forces. This so-called kinematic assumption is common in swarming models, and we discuss it further in Section \ref{sec:discretemodel}.

The first term on the right-hand side of (\ref{eq:locusts}) describes endogenous forces; $\q(r)$ measures the force that locust $j$ exerts on locust $i$. The first term of $\q(r)$ describes attraction, which operates with strength $\F$ over a length scale $L$ and is necessary for aggregation. The second term is repulsive, and operates more strongly and over a shorter length scale in order to prevent collisions. Time and space are scaled so that the repulsive strength and length scale are unity. The second term on the right-hand side of (\ref{eq:locusts}) describes gravity, acting downwards with strength $g$. The last term describes advection of locusts in the direction of the wind with speed $U$. Furthermore, the model assumes a flat impenetrable ground. Since locusts rest and feed while grounded, their motion in that state is negligible compared to their motion in the air. Thus we add to (\ref{eq:locusts}) the stipulation that grounded locusts whose vertical velocity is computed to be negative under (\ref{eq:locusts}) remain stationary.

As mentioned above, for some parameters, (\ref{eq:locusts}) forms a bubble-like shape. This can occur even in the absence of wind, that is, when $U=0$; see Figure \ref{fig:locust}(b). The bubble is crucial, for it allows the swarm to roll in the presence of wind. As discussed in \cite{TopBerLog2008}, states which lack a bubble in the absence of wind do not migrate in the presence of wind. Conditions for bubble formation, even in the equilibrium state arising in the windless model, have not been determined; we will investigate this problem.

Some swarming models adopt a discrete approach -- as in our locust example above -- because of the ready connection to biological observations. A further advantage is that simulation of discrete systems is straightforward, requiring only the integration of ordinary differential equations. However, since biological swarms contain many individuals, the resulting high-dimensional systems of differential equations can be difficult or impossible to analyze. Furthermore, for especially large systems, computation, though straightforward, may become a bottleneck. Continuum models are more amenable to analysis. One well-studied continuum model is that of \cite{BodVel2005}, a partial integrodifferential equation model for a swarm population density $\rho(x,t)$ in one spatial dimension:
\begin{equation}
\label{eq:introeq}
\rho_t + (\rho V)_x = 0, \quad V(x) = \int_{\mathbb{R}} q(x-y) \rho(y)\,dy.
\end{equation}
The density $\rho$ obeys a conservation equation, and $V$ is the velocity field, which is determined via convolution with the antisymmetric pairwise endogenous force $q$, the one-dimensional analog of a social force like the one in (\ref{eq:locusts}).

The general model (\ref{eq:introeq}) displays at least three solution types as identified in~\cite{BodVel2006}. Populations may concentrate to a point, reach a finite steady state, or spread. In~\cite{LevTopBer2009}, we identified conditions on the social interaction force $q$ for each behavior to occur. These conditions map out a ``phase diagram'' dividing parameter space into regions associated with each behavior. Similar phase diagrams arise in a dynamic particle model \cite{DOrChuBer2006} and its continuum analog \cite{ChuDOrMar2007}. Models that break the antisymmetry of $q$ (creating an asymmetric response of organisms to each other) display more complicated phenomena, including traveling swarms \cite{MilYan2008}.

Many studies have sought conditions under which the population concentrates to a point mass. In a one-dimensional domain, collapse occurs when the force $q$ is finite and attractive at short distances \cite{BodVel2006}. The analogous condition in higher dimensions also leads to collapse \cite{BerLau2007,BerLau2009,BerCarLau2009,BerBra2010,BerLauRos2010}. One may also consider the case when the velocity includes an additional term describing an exogenous force,
\begin{equation}
\label{eq:introeq2}
V(x) = \int_{\mathbb{R}} q(x-y) \rho(y)\,dy + f(x).
\end{equation}
In this case, equilibrium solutions consisting of sums of point-masses can be linearly and nonlinearly stable, even for social forces $q$ that are repulsive at short distances \cite{FelRao2010,FelRao2010b,Rao2010}. These results naturally lead to the question of whether a solution can be continued past the time at which a mass concentrates. Early work on a particular generalization of (\ref{eq:introeq}) suggests the answer is yes \cite{HolPut2005,HolPut2006}. For (\ref{eq:introeq}) itself in arbitrary dimension, there is an existence theory beyond the time of concentration \cite{CarDifFig2010}.

Some of the concentration solutions mentioned above are equilibrium solutions. However, there may be classical equilibria as well. For most purely attractive $q$, the only classical steady states are constant in space, as shown via a variational formulation of the steady state problem \cite{BodVel2006}. However, these solutions are non-biological, as they contain infinite mass. There do exist attractive-repulsive $q$ which give rise to compactly-supported classical steady states of finite mass. For instance, in simulations of  (\ref{eq:introeq}), we found classical steady state solutions consisting of compactly supported swarms with jump discontinuities at the edges of the support \cite{LevTopBer2009}. In our current work, we will find equilibria that contain both classical and nonclassical components.

Many of the results reviewed above were obtained by exploiting the underlying gradient flow structure of (\ref{eq:introeq2}). There exists an energy functional
\begin{equation}
W[\rho] = \frac{1}{2} \int_\mathbb{R} \int_\mathbb{R} \rho(x) \rho(y) Q(x-y)\,dx\,dy + \int_\mathbb{R} F(x)\rho(x)\,dx,
\end{equation}
which is minimized under the dynamics. This energy can be interpreted as the continuum analog of the summed pairwise energy of the corresponding discrete (particle) model \cite{Vil2003,BodVel2005,CarDifFig2010}. We will also exploit this energy to find equilibrium solutions and study their stability.

In this paper, we focus on equilibria of swarms and ask the following questions:
\begin{itemize}
\item What sorts of density distributions do swarming systems make? Are they classical or nonclassical?
\item How are the final density distributions reached affected by endogenous interactions, exogenous forces, boundaries, and the interplay of these?
\item How well can discrete and continuum swarming systems approximate each other?
\end{itemize}
To answer these questions, we formulate a general mathematical framework for discrete, interacting swarm members in one spatial dimension, also subject to exogenous forces. We then derive an analogous continuum model and use variational methods to seek minimizers of its energy. This process involves solution of a Fredholm integral equation for the density. For some choices of endogenous forces, we are able to find exact solutions. Perhaps surprisingly, they are not always classical. In particular, they can involve $\delta$-function concentrations of mass at the domain boundary.

The rest of this paper is organized as follows. In Section \ref{sec:formulation}, we create the mathematical framework for our study, and derive conditions for a particular density distribution to be an equilibrium solution, and to be stable to various classes of perturbations.

In Sections \ref{sec:repulsive} and \ref{sec:morse}, we demonstrate different types of swarm equilibria via examples. In Section \ref{sec:repulsive}, we focus on purely repulsive endogenous interactions. We consider a bounded domain with no exogenous forces, a half-line subject to gravitational forces, and an unbounded domain subject to a quadratic exogenous potential, modeling attraction to a light, chemical, or nutrient source. For all three situations, we find exact solutions for swarm equilibria. For the first two examples, these equilibria consist of a density distribution that is classical in the interior of the domain, but contains $\delta$-functions at the boundaries. For the third example, the equilibrium is compactly supported with the density dropping discontinuously to zero at the edge of the support. For all three examples, we compare analytical solutions from the continuum framework to equilibria obtained from numerical simulation of the underlying discrete system. The two agree closely even for small numbers of discrete swarm members. Section \ref{sec:morse} is similar to Section \ref{sec:repulsive}, but we now consider the more complicated case of endogenous interactions that are repulsive on short length scales and attractive over longer ones; such forces are typical for swarming biological organisms.

In Section \ref{sec:locust-ground}, we revisit locust swarms, focusing on their bubble-like morphology as described above, and on the significance of dimensionality. In a one-dimensional model corresponding to a vertical slice of a wide locust swarm under the influence of social interactions and gravity, energy minimizers can reproduce concentrations of locusts on the ground and a group of locusts above the ground, but there cannot be a separation between the two groups. However, a quasi-two-dimensional model accounting for the influence of the swarm's horizontal extent does, in contrast, have minimizers which qualitatively correspond to the biological bubble-like swarms.

\section{Mathematical formulation}
\label{sec:formulation}

\subsection{Discrete model}
\label{sec:discretemodel}

Consider $N$ identical interacting particles (swarm members) in one spatial dimension with positions
$x_i$. Assume that motion is governed by Newton's law, so that acceleration is proportional to the sum of the drag and motive forces. We will focus on the case where the acceleration is negligible and the drag force is proportional to the velocity. This assumption is appropriate when drag forces dominate momentum, commonly known in fluid dynamics as the low Reynolds number or Stokes flow regime. In the swarming literature, the resulting models, which are first-order in time, are known as \emph{kinematic}. Kinematic models have been used in numerous studies of swarming and collective behavior, including \cite{VicCziBen1995,EdeWatGru1998,MogEde1999,GreChaTu2001,CouKraJam2002,GreChaTu2003,MogEdeBen2003,GreCha2004,TopBer2004,TopBerLew2006,TopBerLog2008}).

We now introduce a general model with both endogenous and exogenous forces, as with the locust model (\ref{eq:locusts}). The endogenous forces act between individuals and might include social attraction and repulsion; see~\cite{MogEdeBen2003} for a discussion. For simplicity, we assume that the endogenous forces act in an additive, pairwise manner. We also assume that the forces are symmetric, that is, the force induced by particle $i$ on particle $j$ is the opposite of that induced by particle $j$ on particle $i$. Exogenous forces might include gravity, wind, and taxis towards light or nutrients.

The governing equations take the form 
\begin{subequations}
\label{eq:discretesystem}
\begin{gather}
\frac{dx_i}{dt} = \Vee_i(x_1,\ldots,x_N), \\
\Vee_i(x_1,\ldots,x_N) = \mathop{\sum_{j=1}^{N}}_{j \neq i} m 	q(x_i - x_j) + f(x_i), \quad m = M/N. \label{eq:Vee}
\end{gather}
\end{subequations}
Eventually we will examine the governing equations for a continuum limit of the discrete problem. To this end, we have introduced a \emph{social mass} $m$ which scales the strength of the endogenous forces so as to remain bounded for $N \to \infty$. $M$ is the total social mass of the ensemble. Eq. (\ref{eq:Vee}) defines the velocity rule; $mq$ is the endogenous velocity one particle induces on another, and $f$ is the exogenous velocity. From our assumption of symmetry of endogenous forces, $q$ is odd and in most realistic situations is discontinuous at the origin.

Each force, $f$ and $q$, can be written as the gradient of a potential under the relatively minor assumption of integrability. As pointed out in \cite{MogEdeBen2003}, most of the specific models for mutual interaction forces proposed in the literature satisfy this requirement. Many exogenous forces -- including gravity and common forms of chemotaxis -- do so as well. Under this assumption, we rewrite (\ref{eq:discretesystem}) as a gradient flow,
\begin{equation}
\label{eq:discretegradient1}
\frac{dx_i}{dt} = \Vee_i(x_1,\ldots,x_N) \equiv -\nabla_i \Wee(x_1,\ldots,x_N),
\end{equation}
where the potential $\Wee$ is
\begin{subequations}
\label{eq:discrete_gradient}
\begin{gather}
\Wee(x_1,\ldots,x_N) = \frac{1}{2} \sum_{i = 1}^{N} \mathop{\sum_{j=1}^{N}}_{i \neq j}
m Q(x_i - x_j) +\sum_{k=1}^{N} F(x_k), \label{eq:generic_potential} \\
Q(z) = -\int^z q(s)\,ds,\quad F(z) = -\int^z f(s)\,ds.
\end{gather}
\end{subequations}
The double sum describes the endogenous forces and the single sum describes the exogenous forces. Also,  $Q$ is the mutual interaction potential, which is even, and $F$ is the exogenous potential. The flow described by (\ref{eq:discretegradient1}) will evolve towards minimizers of the energy $\Wee$.

Up to now, we have defined the problem on $\mathbb{R}$. In order to confine the problem to a particular domain $\Omega$, one may use the artifice of letting the exogenous potential $F$ tend to infinity on the complement of $\Omega$.

While this discrete model is convenient from a modeling and simulation standpoint, it is difficult to analyze. Presently, we will derive a continuum analog of (\ref{eq:discretesystem}). This continuum model will allow us to derive equilibrium solutions and determine their stability via the calculus of variations and integral equation methods.

\subsection{Continuum model}
\label{sec:contmodel}

To derive a continuum model, we begin by describing our evolving ensemble of discrete particles with a density function $\rho(x,t)$ equal to a sum of $\delta$-functions. (For brevity, we suppress the $t$ dependence of $\rho$ in  the following discussion.) Our approach here is similar to \cite{BodVel2005}. These $\delta$-functions have strength $m$ and are located at the positions of the particles:
\begin{equation}
\label{eq:deltafuncs}
\rho(x) = \sum_{i=1}^{N} m \delta(x-x_i).
\end{equation}
The total mass is
\begin{equation}
\label{eq:mass_constraint}
M = \int_\Omega \rho(x)\,dx = mN,
\end{equation}
where $\Om$ is the domain of the problem. Using (\ref{eq:deltafuncs}), we write the discrete velocity $\Vee_i(x_1,\ldots,x_N)$ in terms of a continuum velocity $V(x)$. That is, we require $\Vee_i(x_1,\ldots,x_N) = V(x_i)$ where
\begin{equation}
\label{eq:cont_velocity}
V(x) = \int_{\Omega} q(x-y) \rho(y)\,dy + f(x).
\end{equation}
By conservation of mass, the density obeys
\begin{equation}
\label{eq:pde}
\rho_t + (\rho V)_x = 0,
\end{equation}
with no mass flux at the boundary. We now introduce an energy functional $W[\rho]$ which is analogous to the discrete potential $\Wee$ in (\ref{eq:generic_potential}):
\begin{equation}
\label{eq:continuum_energy}
W[\rho] = \frac{1}{2} \int_\Om \int_\Om \rho(x) \rho(y) Q(x-y)\,dx\,dy + \int_\Om F(x)\rho(x)\,dx.
\end{equation}
This expression follows from the discrete potential,
\begin{equation}
W[\rho] = m \Wee(x_1,\ldots,x_N),
\end{equation}
remembering the $\delta$-function definition of the density (\ref{eq:deltafuncs}). The rate of energy dissipation is
\begin{equation}
\frac{dW[\rho]}{dt} = - \int_\Omega  \rho(x) \left\{V(x)\right\}^2\,dx,
\end{equation}
where we assume that there is no mass flux at the boundary of $\Omega$. A consequence of this boundary condition is that under some conditions, mass may concentrate at the boundary of the domain, and we will later see this manifest. Since energy is dissipated, we conclude that stable equilibria correspond to minimizers of $W$.

\begin{figure}
\centerline{\resizebox{\textwidth}{!}{\includegraphics{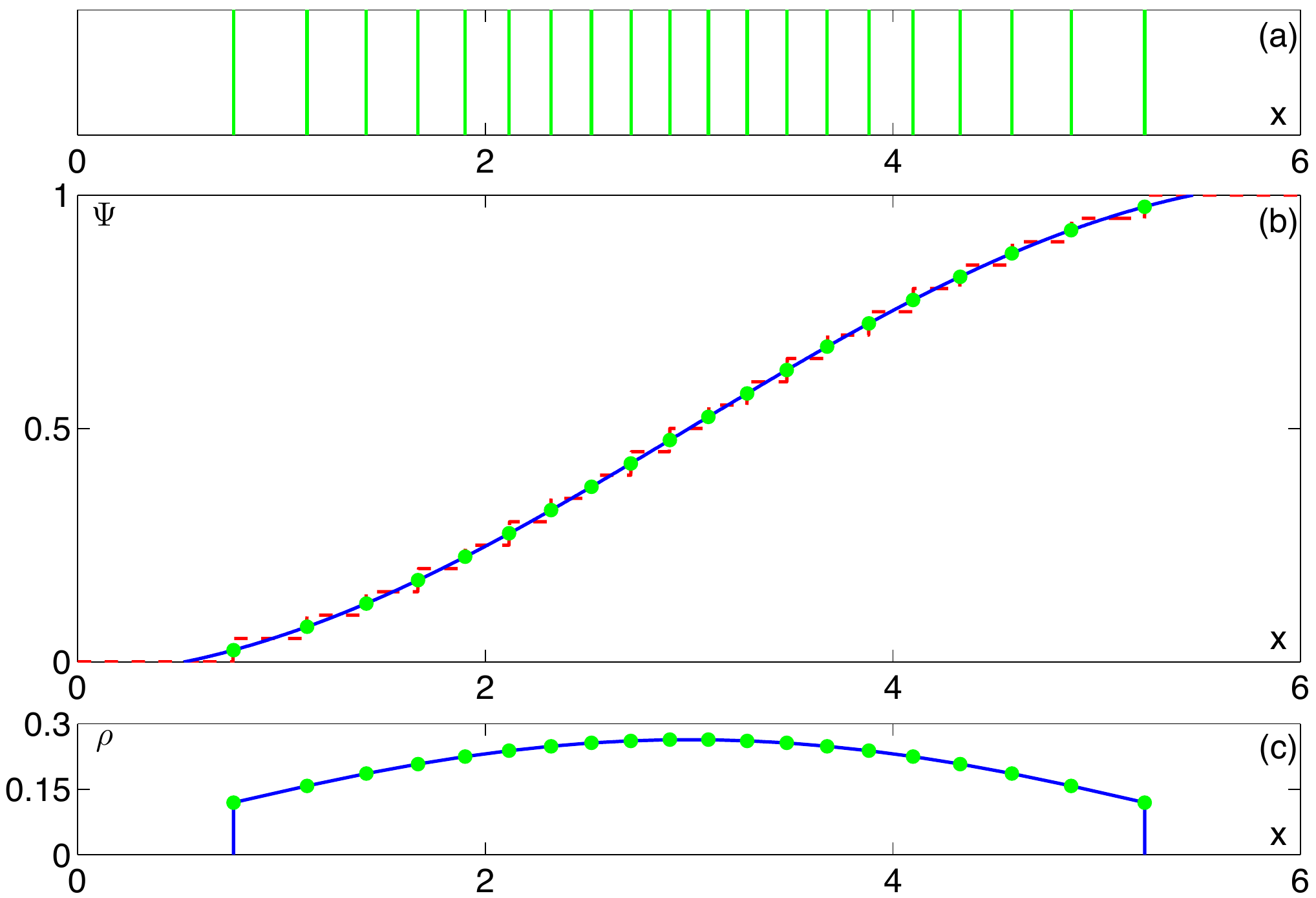}}}
\caption{Schematic correspondence between discrete and continuum systems in one spatial dimension.~(a) Positions of individual swarm members, modeled as a sum of $\delta$-functions per~(\ref{eq:rhodiscrete}).~(b)~Cumulative density distributions. The discrete cumulative density $\Psi_d(x)$ appears as a step function (dotted red curve) per (\ref{eq:Psid}). The continuum cumulative density $\Psi_c(x)$ (solid blue curve) is obtained by interpolating $\Psi_d(x)$ such that the two agree at the positions of the swarm members (green dots).~(c) Continuum density $\rho_c(x)$ corresponding to (a), obtained by differentiating the cumulative density $\Psi_c(x)$ in (b).\label{fig:delta_schematic}}
\end{figure}

Imagine now that the energy (\ref{eq:continuum_energy}) is defined for \emph{any} density distribution $\rho$, not just ensembles of $\delta$-functions. We will find minimizers for the continuous energy (\ref{eq:continuum_energy}) and show that they approximate solutions to the discrete problem in the limit of large $N$. To establish a correspondence between the two frameworks, consider a continuous distribution $\rho_c$ with total mass $M$. Define the cumulative density function
\begin{equation}
\Psi_c(x) = \int_{x_0}^x \rho_c(s)\,ds,
\end{equation}
where the dummy coordinate $x_0$ is taken to the left of the support of $\rho_c$. We seek a discrete approximation of $N$ $\delta$-functions,
\begin{equation}
\label{eq:rhodiscrete}
\rho_d(x) = \sum_{i=1}^{N} m \delta(x-x_i).
\end{equation}
The associated cumulative density function $\Psi_d$ is
\begin{equation}
\label{eq:Psid}
\Psi_d(x) = 
\begin{cases}
0 & x < x_1 \\
m[1/2 + (i-1)] & x = x_i,\quad i = 1,\ldots,N \\
im & x_i < x < x_{i+1}, \quad i = 1,\ldots,N-1\\
M & x > x_N,
\end{cases}
\end{equation}
where we have used the convention that integrating up to a $\delta$-function yields half the mass of integrating through it. To establish our correspondence, we require that  $\Psi_c(x_i) = \Psi_d(x_i)$, which in turn determines the particle positions $x_i$. As $N \to \infty$ for fixed $M$, this step function $\Psi_d$ converges uniformly to $\Psi_c$. The correspondence goes in the opposite direction as well. We can begin with an ensemble of $\delta$-functions $\rho_d$ placed at the positions of discrete swarm members, as shown in Figure \ref{fig:delta_schematic}(a). We can find the corresponding cumulative density $\Psi_d$ via (\ref{eq:Psid}) and interpolate to construct the continuum cumulative density $\Psi_c$. The functions $\Psi_d$ and $\Psi_c$ are shown as the dotted red step function and the blue curve, respectively, in Figure \ref{fig:delta_schematic}(b). We may then differentiate to find an approximation $\rho_c$, as shown in Figure \ref{fig:delta_schematic}(c). We use this correspondence in Sections \ref{sec:repulsive} through \ref{sec:locust-ground} to compare analytical results for the continuum system (\ref{eq:pde}) with numerical simulations of the discrete system (\ref{eq:discretesystem}).

We close this subsection by reiterating why we have made a correspondence between the discrete and continuum systems. We use the continuous framework to find equilibrium solutions analytically via variational and integral equation methods. The correspondence above allows a direct comparison to numerical simulation of the discrete system.

\subsection{Local and global minimizers}
\label{sec:minimizers}

We use a variational calculation to determine conditions for a density distribution to be a minimizer of $W$. Our starting point is the energy functional (\ref{eq:continuum_energy}) subject to the mass constraint (\ref{eq:mass_constraint}). Let
\begin{equation}
\label{eq:expand_rho}
\rho(x) =\brho + \epsilon{\tilde \rho}(x).
\end{equation}
Here $\brho$ is an equilibrium solution of mass $M$ and $\epsilon \trho$ is a small perturbation of zero mass, so we have
\begin{subequations}
\begin{gather}
\int_\Omega \brho (x)\,dx = M \label{eq:mass1}, \\ 
\int_\Omega \trho(x)\,dx = 0.
\end{gather}
\end{subequations}
Inspired by biological observations of swarms, we focus on equilibria with finite extent and take the support of $\brho$ to be a finite subset of the domain $\Omega$. We refer to the support of $\brho$ as $\Op$. This assumption, combined with the fact that the density is nonnegative, restricts the perturbation $\trho$ to be nonnegative on $\Op^c$, the complement of $\Op$.

We write
\begin{equation}
\label{eq:expand_W}
W[\rho] =W[\brho] + \epsilon W_1[\brho,\trho] + \epsilon^2 W_2[\trho,\trho],
\end{equation}
where $W_1$ and $W_2$ are the first and second variations respectively. This expression is exact because $W$ is quadratic in $\rho$ (see eq.~\ref{eq:continuum_energy}). We analyze these variations to determine necessary and sufficient conditions for a candidate solution $\brho$ to be a minimizer of $W$.

Our strategy is to consider two classes of perturbations $\trho$. First, we consider perturbations whose support lies in $\Op$. In order for $\brho$ to be extremal, $W_1$ must vanish. For it to be a minimizer, $W_2$ must be positive. Since (\ref{eq:expand_W}) is exact, $W_2 > 0$ guarantees that $\brho$ is both a local and global minimum with respect to this first class of perturbations.

The second class of perturbations we consider (of which the first class is a subset) consists of perturbations on the entire domain $\Om$. As mentioned above, these perturbations must be nonnegative in $\Op^c$ in order to maintain positivity of $\rho$. A necessary condition for $\brho$ to be a local minimizer is that $W_1 \geq 0$ for this class of perturbations. If in addition $W_2$ remains positive for this larger class of perturbations, then $\brho$ is a global minimizer as well.

We derive the first variation $W_1$ by substituting (\ref{eq:expand_rho}) into (\ref{eq:expand_W}) and expanding to first order in $\epsilon$, which yields 
\begin{equation}
\label{eq:first_variation}
W_1[\brho,\trho] =  \int_\Omega \trho \left [ \int_\Omega Q(x-y) \brho (y)\,dy+F(x)  \right ]\,dx .
\end{equation}
Consider the first class of perturbations, whose support lies in $\Op$. As the perturbation $\trho$ is arbitrary and of zero mass, for the first variation $W_1$ to vanish, it must be true that
\begin{equation}
\label{eq:lagrange}
\int_\Op Q(x-y) \brho(y)\,dy + F(x) = \lambda \quad \textrm{for $x \in \Op$.}
\end{equation}
The same result can be found by a Lagrange multiplier argument including the constant mass as a constraint. The multiplier $\lambda$ has a physical interpretation; it is the energy per unit mass an additional test mass would feel due to the exogenous potential and the interaction with $\brho$. 

Thus far, we have shown that a necessary condition for $\brho$ to be an equilibrium solution is that it satisfies the Fredholm integral equation of the first kind for the nonnegative density $\brho$,
\begin{equation}
{\cal I}[\brho(x)] = \lambda - F(x), \quad {\cal I}[\brho(x)] \equiv \int_\Op Q(x-y)  \brho(y)\,dy,
\label{eq:FIE}
\end{equation}
as well as the mass constraint (\ref{eq:mass1}).

In order for $\brho$ to be a minimizer with respect to the first class of perturbations, the second variation must be positive. Substituting (\ref{eq:expand_rho}) into (\ref{eq:expand_W}) yields
\begin{equation}
\label{eq:second_variation}
W_2[\trho,\trho] = \int_\Op  \int_{\Op}  Q(x-y) \trho(x) \trho(y)\,dx\,dy.
\end{equation}
The sign of $W_2$ can be assessed in a number of ways.

We first derive a sufficient condition on the Fourier transform of $Q$ for $W_2$ to be nonnegative for the first class of perturbations. Define the Fourier transform
\begin{equation}
\widehat{Q}(k) = \int_{-\infty}^{\infty} Q(x)\exp{-ikx}\,dx.
\end{equation}
Then we have
\begin{subequations}
\begin{eqnarray}
W_2[\trho,\trho] & = & \int_\Op  \int_{\Op}  Q(x-y) \trho(x) \trho(y)\,dx\,dy,\\
& = & \int_{-\infty}^{\infty}  \int_{-\infty}^{\infty}  Q(x-y) \trho(x) \trho(y)\,dx\,dy, \\
& = & \int_{-\infty}^{\infty} \trho(x) \left[ Q(x) * \trho(x) \right]\,dx, \\
& = & \frac{1}{2\pi} \int_{-\infty}^{\infty} | \widehat{\trho}(k)|^2 \widehat{Q}(k)~dk.
\end{eqnarray}
\end{subequations}
We have used the fact that $\trho$ is compactly supported to extend the range of integration to infinity. We have also used the convolution theorem, Parseval's theorem, and the fact that $\trho$ is real. We see, then, that $\widehat{Q}(k) > 0$ is a sufficient condition for $W_2 > 0$ (assuming a nontrivial perturbation). As shown in \cite{LevTopBer2009}, this condition is actually equivalent to that for the linear stability of a constant density state in the absence of exogenous forces.

A necessary and sufficient condition for $W_2 > 0$ for the first class of perturbations comes from considering the spectrum of $\cal{I}$ in $\Op$. Note that (\ref{eq:second_variation}) may be written as
\begin{equation}
W_2[\trho,\trho] = \int_\Op   {\cal I}[\trho]\trho(x)\, dx \equiv \langle \trho,  {\cal I}[\trho] \rangle,
\end{equation}
where the angle brackets denote the usual $L^2$ inner product on $\Op$. If the eigenvalues of the integral operator $\cal{I}$ are positive, then $W_2 > 0$ (again assuming a nontrivial perturbation).

We now turn to the second class of perturbations, which have support in $\Om$ and which are positive in $\Op^c$. To analyze these perturbations, we extend the definition of the constant $\lambda$ to a function $\Lambda(x)$ that is defined over all of $\Omega$. We set
\begin{equation}
\label{eq:Lambda}
\Lambda(x) \equiv \int_\Op Q(x-y)  \brho(y)\,dy +F(x).
\end{equation}
Trivially from (\ref{eq:lagrange}), $\Lambda(x) = \lambda$ for $x \in \Op$. We now rewrite the first variation as
\begin{equation}
\label{eq:W1Lambda}
W_1[\brho,\trho] = \int_\Om \trho(x) \Lambda(x)\,dx,
\end{equation}
directly from (\ref{eq:first_variation}). Remembering that $\trho \geq 0$ in $\Op^c$ and that $\trho$ has zero mass in $\Om$, we see that a necessary and sufficient condition for $W_1 \geq 0$ is $\Lambda(x) \geq \lambda$ in $\Op^c$, that is
\begin{equation}
\label{eq:stable}
\Lambda(x) \equiv \int_\Op Q(x-y)  \brho(y)\,dy +F(x) \geq \lambda \quad \textrm{for $x \in \Op^c$.}
\end{equation}
Physically, this guarantees that a parcel of mass transported from $\Op$ to its complement increases the total energy.

If we wish for $\brho$ to be a minimizer with respect to the second class of perturbations, it suffices for $W_2 > 0$ (for example, by having $\widehat{Q}(k) > 0$). However, this condition is not necessary. The necessary condition is that $W_1[\brho,\trho] + W_2[\trho,\trho] > 0$ for nontrivial perturbations, which follows from (\ref{eq:expand_W}) being exact.

To summarize, we have obtained the following results:
\begin{itemize}
\item{Equilibrium solutions $\brho$ satisfy the Fredholm integral equation (\ref{eq:FIE}) and the mass constraint (\ref{eq:mass1}).}
\item{The solution $\brho$ is a local and global minimizer with respect to the first class of perturbations (those with support in $\Op$) if $W_2$ in (\ref{eq:second_variation}) is positive.}
\item{The solution $\brho$ is a local minimizer with respect to the second (more general zero-mass) class of perturbations if $\brho$ satisfies (\ref{eq:stable}). If in addition $W_2$ is positive for these perturbations, then $\brho$ is a global minimizer as well.}
\end{itemize}
In practice, we solve the integral equation (\ref{eq:FIE}) to find candidate solutions. Then, we compute $\Lambda(x)$ to determine whether $\brho$ is a local minimizer. Finally, when possible, we show the positivity of $W_2$ to guarantee that $\brho$ is a global minimizer.

\subsection{Swarm minimizers}

As the continuum limit replaces individual particles with a density, we need to make sure the continuum problem inherits a physical interpretation for the underlying problem. If we think about perturbing an equilibrium configuration, we note that mass cannot ``tunnel'' between disjoint components of the solution. As such we define the
concept of a multi-component swarm equilibrium. Suppose the swarm's support can be divided into a set of $m$ disjoint, closed, connected components $\{\Om_i\}$, that is 
\begin{equation}
\Op =\Om_1 \cup \Om_2 \cup  \cdots \cup \Om_m, \qquad \Om_i \cap \Om_j = \varnothing, \quad i \ne j.
\end{equation}
We define a swarm equilibrium as a configuration in which each individual swarm component is in equilibrium,
 \begin{equation}
 \int_{\Om_i}
Q(x-y)  \rho(y)~dy + F(x) 
= \lambda_i \qquad {\rm for} \quad i = 1, \ldots, m .
\end{equation}
We can still define $\Lambda(x)$ in $\Om$
\begin{equation}
 \Lambda(x)
 =  {\cal I}[\brho(x)] +F(x)  =
 \int_\Op
Q(x-y)  \brho(y)~dy +F(x), 
\end{equation}
but now $\Lambda(x)=\lambda_i$ in $\Om_i$. We can now define a swarm minimizer. We say a swarm equilibrium is a swarm minimizer if $\Lambda(x) \ge \lambda_i$ for some neighborhood of each component $\Om_i$ of the swarm.
In practice this means that the swarm is an energy minimizer for infinitesimal redistributions of mass in the neighborhood of each component. This might also be called a Lagrangian minimizer in the sense that the equilibrium is a minimizer with respect to infinitesimal Lagrangian deformations of the distributions.

It is crucial to  note that even if a solution $\brho$ is a global minimizer, other multi-component swarm minimizers may still exist. These solutions are local minimizers and consequently a global minimizer may not be a global attractor under the dynamics of~(\ref{eq:pde}).


\section{Examples with a repulsive social force}
\label{sec:repulsive}

\begin{figure}
\centerline{\resizebox{\textwidth}{!}{\includegraphics{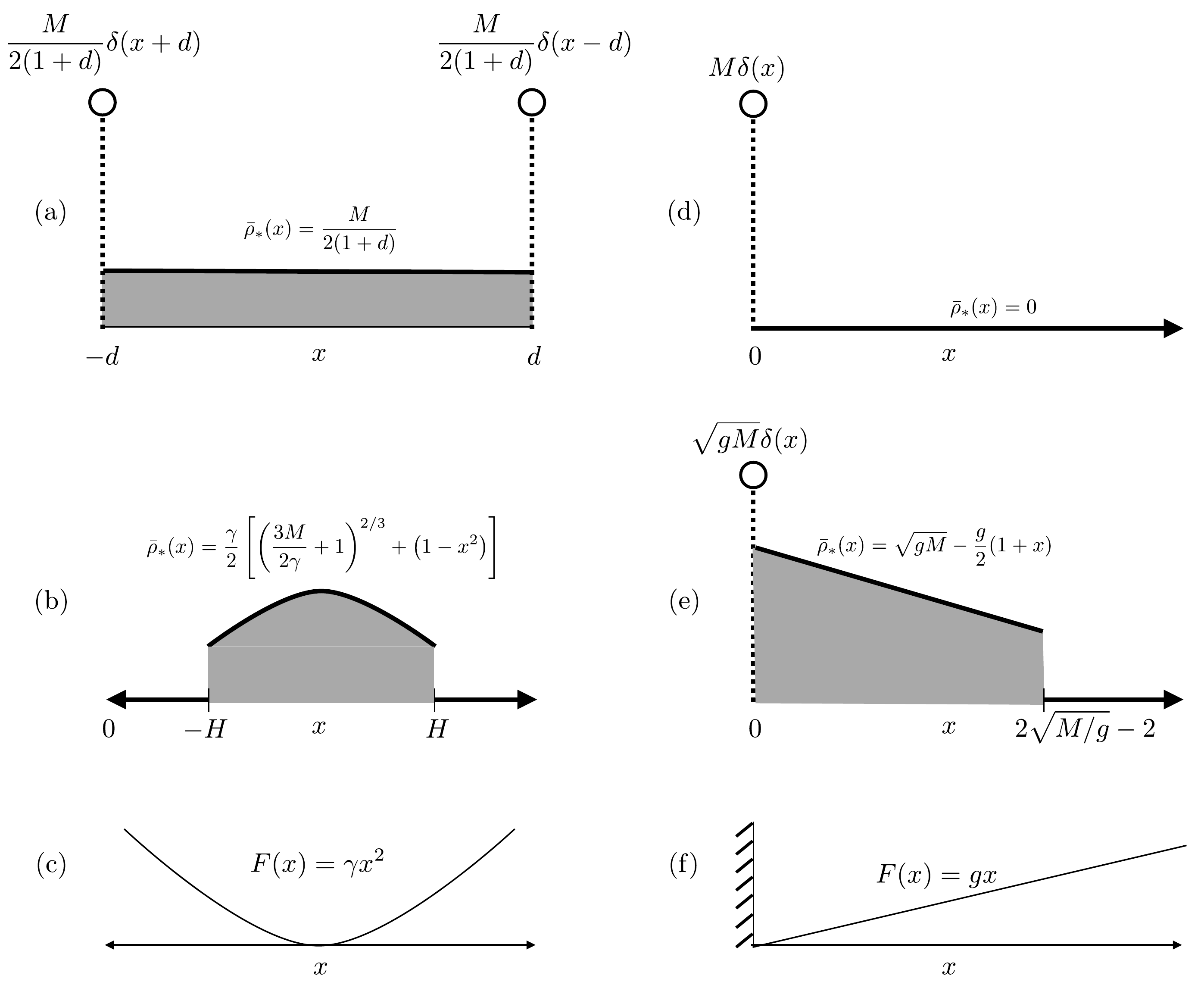}}}
\caption{Schematic solutions of the minimization problem defined by (\ref{eq:FIE}) and (\ref{eq:mass1}) when the potential $Q(x)$ is chosen to be the purely repulsive Laplace potential (\ref{eq:laplace}). The total mass is $M$. Minimizers $\brho(x)$ are shown in (a,b,d,e) and two exogenous potentials $F(x)$ are shown in (c,f).~(a)~The minimizer for the bounded domain $\Om = [-d,d]$ with $F(x) = 0$ is a constant internal density distribution with $\delta$-functions at the boundary. See Section \ref{sec:bounded}.~(b)~The minimizer for the unbounded domain $\Om = \mathbb{R}$ with an exogenous quadratic potential $F(x)$ of strength $\gamma$ is a compactly supported, downward-facing parabola section. See Section \ref{sec:quad}.~(c)~The quadratic potential $F(x)$ corresponding to (b).~(d)~The minimizer for the half-line $\Om = [0,\infty]$ with an exogenous gravitational potential $F(x) = gx$ is a $\delta$-function at the origin for sufficiently large gravity $g$.~(e)~Like (d), but for smaller $g$, in which case the minimizer is a \mbox{$\delta$-function} at the origin, is linear for a finite interval connected to the origin, and drops discontinuously to zero outside of that interval. See Section \ref{sec:grav}.~(f)~The gravitational potential $F(x)$ corresponding to (d,e). \label{fig:repulsion_schematic}}
\end{figure}

In this section we discuss  the minimization problem formulated in Section \ref{sec:formulation}. It is helpful for expository purposes to make a concrete choice for the interaction potential $Q$. As previously mentioned, in many physical, chemical, and biological applications, the pairwise potential $Q$ is symmetric. Additionally, repulsion dominates at short distances (to prevent collisions) and the interaction strength approaches zero for very long distances. A common choice for $Q$ is the Morse potential with parameters chosen to describe long-range attraction and short-range repulsion \cite{MogEdeBen2003}. For the remainder of this section, we consider a simpler example where $Q$ is the Laplace distribution
\begin{equation}
Q(x) =  \exp{-|x|}, \label{eq:laplace}
\end{equation}
which represents repulsion with strength decaying exponentially in space.

When there is no exogenous potential, $F(x) = 0$, and when the domain is infinite, \emph{e.g.}, $\Om= \mathbb{R}$, the swarm will spread without bound. The solutions asymptotically approach the Barenblatt solution to the porous medium equation as shown in \cite{LevTopBer2009}. However, when the domain $\Om$ is bounded or when there is a well in the exogenous potential, bounded swarms are observed both analytically and numerically, as we will show. Figure \ref{fig:repulsion_schematic} shows solutions $\brho(x)$ for three cases: a bounded domain with no exogenous potential, a gravitational potential on a semi-infinite domain, and a quadratic potential well on an infinite domain. In each case, a bounded swarm solution is observed but the solutions are not necessarily continuous and can even contain $\delta$-function concentrations at the boundaries. 

We discuss these three example cases in detail later in this section. First, we will formulate the minimization problem for the case of the Laplace potential. We will attempt to solve the problem classically; when the solution has compact support contained within the domain we find solutions that are continuous within the support and may have jump discontinuities at the boundary of the support. However, when the boundary of the support coincides with the boundary of the domain, the classical solution may break down and it is necessary to include a distributional component in the solution. We also formulate explicit conditions for the solutions to be global minimizers. We then apply these results to the three examples mentioned above.

\subsection{Classical solutions to the integral equation}
\label{sec:classical}

Recall that for $\brho$ to be a steady solution, it must satisfy the integral equation (\ref{eq:FIE}) subject to the mass constraint (\ref{eq:mass1}).
For $\brho(x)$ to be a local minimizer, it must also satisfy (\ref{eq:stable}),
\begin{equation}
\label{eq:stable2}
\Lambda(x) \equiv \int_\Op Q(x-y)  \brho(y)\,dy +F(x)  \geq \lambda.
\end{equation}
Finally, recall that for a solution $\brho$ to be a global minimizer, the second variation (\ref{eq:second_variation}) must be positive. We saw that if $\widehat{Q}(k) > 0$, this is guaranteed. For (\ref{eq:laplace}), $\widehat{Q}(k) = 2/(1+k^2)$ and so for the remainder of this section, we are able to ignore the issue of~$W_2$. Any local minimizer that we find will be a global minimizer.

Additionally, for the remainder of this section, we restrict our attention to cases where the support of the solution $\Op$ is a single interval in $\Om$; in other words, the minimizing solution has a connected support. The reason that we are able to make this restriction follows from the notion of swarm minimization, discussed above. In fact, we can show that there are no multi-component swarm minimizers for the Laplace potential as long as the exogenous potential $F(x)$ is convex, that is, $F^{\prime \prime}(x) \geq 0$ on $\Om$. To see this, assume we have a swarm minimizer with a at least two disjoint components. Consider $\Lambda(x)$ in the gap between two components so that $x \in \Op^c$. We differentiate $\Lambda(x)$ twice to obtain
\begin{equation}
\label{eq:singlecomponent}
\Lambda^{\prime \prime}(x) \equiv \int_\Op Q^{\prime \prime}(x-y)  \brho(y)\,dy +F^{\prime \prime}(x).
\end{equation}
Note that $Q^{\prime \prime}(x-y) > 0$ as $x \neq y$ in $\Op^c$.  $F^{\prime \prime}(x) \geq 0$ by assumption. Consequently, $\Lambda^{\prime \prime}(x) > 0$ in $\Op^c$ and so $\Lambda(x)$ is convex upwards in the gap. Also, $\Lambda(x) = \lambda_i$ at the endpoints of the gap. We conclude from the convexity that $\Lambda(x)$ must be less than $\lambda_i$ near one of the endpoints. This violates the condition of swarm minimization from the previous section, and hence the solution is not a swarm minimizer. Since swarm minimization is a necessary condition for global minimization, we now, as discussed, restrict attention to single-component solutions.

For concreteness, assume the support of the solution $\brho$ is $\Op = [\alpha,\beta]$. We transform the integral equation (\ref{eq:FIE}) for $\brho$ into a differential equation by noting that $Q(x)$ is the  Green's function of the differential operator $\mathcal{L} \equiv  \partial_{xx} - 1$, so that $\mathcal{L} Q(x) = -2 \delta(x)$, where $\delta(x)$ is the Dirac $\delta$-function. Applying  $\mathcal{L}$ to both sides of (\ref{eq:FIE}) yields the Dirac $\delta$-function  under the integral. Integration and the sifting property of the $\delta$ function lead to
\begin{equation}
\brho(x) = \brho_*(x) \equiv -\frac{1}{2} {\mathcal L} [\lambda - F(x)] = \frac{\lambda}{2}-\frac{1}{2}\left [F-  F''(x) \right ] 
 \quad {\rm in} \ \Op.
\label{eq:FIE-sol}
\end{equation}

Equation (\ref{eq:FIE-sol}) is a necessary condition on a solution ${\bar \rho}(x)$ but not sufficient. In fact, to verify the candidate solution (\ref{eq:FIE-sol}), we must substitute back into the governing integral equation (\ref{eq:FIE}). We find 
\begin{subequations}
\begin{eqnarray}
{\cal I}[\brho_*(x)] &\equiv&
 \int_\Omega
e^{-|x-y|}  \left \{
 \frac{\lambda}{2}-\frac{1}{2}\left [F(y)-  F''(y) \right ]  \right \}
dy, \\
&=& \lambda - F(x)
+\frac{e^{\alpha-x}}{2}[F(\alpha)-F'(\alpha)-\lambda] \\
& & \mbox{}
+\frac{e^{x-\beta}}{2}[F(\beta)+F'(\beta)-\lambda], \nonumber
\end{eqnarray}
\end{subequations}
where the terms other than $ \lambda - F(x)$ must vanish for (\ref {eq:FIE-sol}) to be a solution. These error terms are spanned by $\{e^x,e^{-x}\}$ which is the null space of ${\mathcal L}$. The constraint that these terms must vanish leads to the conditions
\begin{equation}
F(\alpha)-F'(\alpha)=\lambda,  \qquad F(\beta)+F'(\beta)=\lambda .
\end{equation}
For a given $F(x)$, both these conditions will only be satisfied for particular choices of $\alpha$ and $\beta$. The allowed $\alpha$ and $\beta$ then specify $\lambda$ which in turn determines the mass $M$ through (\ref{eq:mass1}). As we discuss below, the solution derived here is sometimes, but not always, a true minimizer of the energy $W$. 

\subsection{Functional minimization viewpoint and nonclassical solutions}
\label{sec:funcmin}

What we have presented above is the classical view of the solution of (\ref{eq:FIE}). The difficulty with this calculation is that we expect on physical grounds that for every mass $M$  and interval $\alpha < x < \beta$ there should be a minimizer. It is easy to see that if $F$ and $Q$ are bounded from below,  the energy $W$ is bounded from below also. Let 
\begin{equation}
F_{min}= \min_{x \in \Omega} F(x), \qquad Q_{min}= \min_{x \ge 0}  Q(x).
\end{equation}
Then directly from (\ref{eq:continuum_energy}),
\begin{equation}
W \ge \frac{1}{2} M^2 Q_{min} +M F_{min}.
\end{equation}
Since $W$ is bounded from below, solutions to the minimization problem exist in the space of measure valued functions \cite{Vil2003}. Sometimes these solutions are not classical; when there is finite attraction at small distances, mass can concentrate in a $\delta$-function \cite{BodVel2006,CarDifFig2010}. For the more biologically relevant case of repulsion at short scales, in free space and in the absence of an external potential, solutions are classical for all time \cite{BodVel2006}.

\subsubsection{Absence of $\delta$-concentrations interior to a domain}
\label{sec:absence}

We now show that (under sufficient hypotheses) a $\delta$-function cannot occur in the interior of $\Op$. Though we have restricted ourselves (above) to considering the Laplace potential~(\ref{eq:laplace}), this is true more generally.

In fact, suppose that at a point $x_0$ in the interior of $\Om$,
\begin{itemize}
\item the exogenous potential $F(x)$ is twice-differentiable in a neighborhood of $x_0$,
\item $Q(x)$ is twice-differentiable for $x>0$,
\item $Q(x)$ has nonzero repulsion at short distances, that is,
\begin{equation}
\lim_{x \to 0^\pm} q(x) = \pm K.
\end{equation}
\end{itemize}
Then a minimizer does not contain a $\delta$-function at $x=x_0$.

To see this, we will compare the energy of a density containing a $\delta$-function at $x=x_0$ to one where the $\delta$-function has been replaced by a narrow, unit mass top-hat distribution $\delta_\varepsilon(x-x_0)$,
\begin{equation}
\delta_\varepsilon(x) =
\begin{cases}
1/(2\varepsilon) & |x| < \varepsilon \\
0 & |x| \geq \varepsilon,
\end{cases}
\end{equation}
and show that the energy is reduced by $\order(\epsilon)$.

Consider a candidate distribution with a $\delta$-function of mass $m$ at $x_0$, $\rho(x) = \rho_0(x) + m \delta(x-x_0)$. Define the change of energy
\begin{equation}
\Delta W = W[\brho_0 + m \delta_\varepsilon(x-x_0)] - W[\brho_0 + m \delta(x-x_0)].
\end{equation}
By direct calculation using (\ref{eq:continuum_energy}),
\begin{eqnarray}
\Delta W & = & \frac{m^2}{2} \int_\Om \int_\Om \delta_\varepsilon(x-x_0) Q(x-y) \delta_\varepsilon(y-x_0) \,dx\,dy - m^2 Q(0)/2  \label{eq:nodeltas} \\
& & \mbox{}+ m \int_\Om \int_\Om \rho_0(y)  Q(x-y) [\delta_\varepsilon(x-x_0) - \delta(x-x_0) ] \,dx\,dy \nonumber \\
& & \mbox{}+ m \int_\Om F(x) \left[\delta_\varepsilon(x-x_0) - \delta (x-x_0) \right]\,dx.  \nonumber
\end{eqnarray}
Note that near $x=0$, $Q(x)=Q(0)-K|x|+\order(x^2)$ and $F(x)=F(x_0)+F^\prime(x_0)(x-x_0)+\order(x^2)$.  Expanding to $\order(\epsilon^2)$, only the first term of (\ref{eq:nodeltas}) persists, yielding
\begin{equation}
\Delta W = -\frac{Km^2}{2}\frac{1}{4\epsilon^2} \int_{-\varepsilon}^{\varepsilon} \int_{-\varepsilon}^{\varepsilon} |x-y|\,dx\,dy + \order(\varepsilon^2) = -\frac{1}{3}\varepsilon m^2 K + \order(\varepsilon^2). \label{eq:nodeltas2}
\end{equation}
For $K>0$, corresponding to nonzero repulsion at short distances, $\Delta W < 0$, indicating that the energy is reduced by replacing the $\delta$-function with a narrow top-hat.

We conclude that $\delta$-functions may not occur in the interior of $\Om$ under the assumed conditions. However, the above reasoning breaks down for $\delta$-functions on the domain boundary, where they may not be replaced by a narrow, symmetric top-hat distribution.

\subsubsection{$\delta$-concentrations on a domain boundary}

Based on the result above, we introduce a candidate solution
\begin{equation}
\brho(x) \equiv \brho_*(x) + A \delta(x-\alpha) +B\delta(x-\beta)
 \quad {\rm in} \ \Omega,
\label{eq:FIE-sol2}
\end{equation}
where $A$ and $B$ are to be determined. The classical solution $\brho_*$ is supplemented with $\delta$-functions at the boundary of $\Op$. We will show, in agreement with the calculation above, that $A$ and $B$ must vanish unless the boundary of the support, $\Op$, coincides with the boundary of the domain, $\Om$. 

We verify the new candidate solution (\ref{eq:FIE-sol2}) by substituting back into the governing equation (\ref{eq:FIE}). We find
\begin{subequations}
\begin{eqnarray}
{\cal I}[\brho(x)] &\equiv&
 \int_\Omega
e^{-|x-y|}  \left \{ {\bar \rho(y)} + A \delta(y-\alpha) +B\delta(y-\beta)
   \right \}\,dy, \\
&=& \lambda - F(x)
+\frac{e^{\alpha-x}}{2}[2A+F(\alpha)-F'(\alpha)-\lambda] \\
& & \mbox{}+\frac{e^{x-\beta}}{2}[2B+F(\beta)+F'(\beta)-\lambda], \nonumber
\end{eqnarray}
\end{subequations}
where the terms other than $ \lambda - F(x)$ must vanish for (\ref {eq:FIE-sol2}) to be a solution. This constraint leads to the following conditions on $A$ and $B$, the coefficients of the nonclassical parts of the solution:
\begin{equation}
A= \frac{\lambda -F(\alpha)+F'(\alpha) }{2}, \quad
B= \frac{\lambda -F(\beta)-F'(\beta)}{2}. 
\label{eq:AB}
\end{equation}
Because  $\lambda$ is undetermined as of yet, we can find a solution for any mass $M$.  Substituting the solution $\brho(x)$ into the mass constraint (\ref{eq:mass1}) yields
\begin{subequations}
\begin{eqnarray}
M&=& \int_\Op \brho(x)\,dx, \\
&=& A+B+\lambda \frac{(\beta -\alpha)}{2} 
-\frac{1}{2} \int_\Op F(x) ~ dx + \frac{1}{2}[ F'(\beta)-F'(\alpha)],\\
&=& \lambda \left ( 1 +\frac{(\beta -\alpha)}{2} \right )
-\frac{1}{2} \left [ F(\alpha)+F(\beta)+ \int_\Op F(x)\,dx \right ].
\end{eqnarray}
\end{subequations}
Solving for $\lambda$ in terms of $M$ yields
\begin{equation}
 \lambda = \frac{2 M + F(\alpha)+F(\beta)+ \int_\Op F(x) ~ dx } {2+\beta - \alpha}.
 \label{eq:Mform}
\end{equation}

We've shown that for any $\Op = [\alpha,\beta]$ and any mass $M$ we can find a solution to (\ref{eq:FIE}) with $\brho(x)$ smooth in the interior and with a concentration at the endpoints. However, we haven't yet addressed the issue of $\brho(x)$ being non-negative, nor have we considered whether it is a minimizer.

We next consider whether the extremal solution $\brho$ is a minimizer, which involves the study of (\ref{eq:stable2}). We present a differential operator method that allows us to compute $\Lambda(x)$ and deduce sufficient conditions for $\brho$ to be a minimizer.

We start by factoring the differential operator $\mathcal{L} \equiv  \partial_{xx} - 1 = \cDp \cDm = \cDm \cDp $ 
where $ \cDpm = \partial_x \pm 1$. Applying these operators to the interaction potential $Q$, we see that
\begin{equation}
\cDp Q(x-y)=
\begin{cases}
2e^{x-y} &  y>x \cr
0 &  y < x,
\end{cases}
\qquad
\cDm Q(x-y)=
\begin{cases}
0 &  y>x \cr
-2e^{y-x}& y<x.
\end{cases}
\end{equation}
Substituting $\brho$ in (\ref{eq:FIE-sol2}) into our definition of $\Lambda(x)$ in (\ref{eq:stable2}) yields
\begin{equation}
\int_\Op Q(x-y)  \brho(y)\,dy + AQ(x-\alpha) + BQ(x-\beta)+ F(x) = \Lambda(x). 
\label{eq:one}
\end{equation}
Now consider applying $\cDm$ to (\ref{eq:one}) at a point $x$ in $\Op$. We see  that 
\begin{equation}
- 2 \int_\alpha^x e^{(y-x)}  \brho(y)~dy 
 -2 A e^{\alpha-x}+ \cDm [F(x)] 
= \cDm [ \lambda],   
\end{equation}
where we've used the fact that $\Lambda(x) =\lambda$ in $\Op$.
If we let   $x=\alpha+z$ and let $z$ decrease to zero, the integral term vanishes and 
\begin{equation}
 -2 A + F'(\alpha) - F(\alpha)  
= - \lambda   .
\end{equation}
Solving for $A$ yields the first half of (\ref{eq:AB}). A similar argument near $x=\beta$ yields the value of $B$.

Assuming $\alpha$ does not coincide with an endpoint of $\Om$, we now consider the region $x<\alpha$, which is to the left of the support $\Op$. Again, applying $\cDm$ to (\ref{eq:FIE}) simplifies the equation; we can check that both the integral term and the contribution from the $\delta$-functions are annihilated by this operator, from which we deduce that
\begin{equation}
\cDm [F(x)- \Lambda(x)] =0 \qquad  \Rightarrow \qquad F(x)- \Lambda(x) = Ce^x,
\end{equation}
where $C$ is an unknown constant. A quick check shows that if $F(x)$ is continuous, then  $\Lambda(x)$ is continuous at the endpoints of $\Op$ so that $\Lambda(\alpha)=\lambda$. This in turn determines $C$, yielding
\begin{equation}
\label{eq:lambdaleft}
\Lambda(x) = F(x) +[\lambda-F(\alpha)]e^{x-\alpha} \qquad {\rm for} \quad x \leq \alpha.
\end{equation}
A similar argument near $x=\beta$ yields 
\begin{equation}
\label{eq:lambdaright}
\Lambda(x) = F(x) +[\lambda-F(\beta)]e^{\beta -x} \qquad {\rm for} \quad x \geq \beta.
\end{equation}
As discussed in Section \ref{sec:minimizers}, for $\brho(x)$ to be a minimizer we wish for $\Lambda(x) \geq \lambda$ for $x \geq \beta$ and $x \leq \alpha$. 
A little algebra shows that this  is equivalent to
\begin{subequations}
\label{eq:mincon}
\begin{gather}
\frac{F(x)e^{-x} - F(\alpha)e^{-\alpha}}{e^{-x}-e^{-\alpha}} \geq  \lambda \quad \textrm{for $x \leq \alpha$}, \label{eq:mincona} \\
\quad \frac{F(x)e^x - F(\beta)e^\beta}{e^x-e^\beta} \geq \lambda \quad \textrm{for $x \geq \beta$}. \label{eq:minconb}
\end{gather}
\end{subequations}
If $\alpha$ and $\beta$ are both strictly inside $\Om$, then (\ref{eq:mincon}) constitutes sufficient conditions for the extremal solution $\brho$ to be a global minimizer (recalling that $W_2 > 0$). We may also derive a necessary condition at the endpoints of the support from (\ref{eq:mincon}). As $x$ increases to $\alpha$, we may apply L'H\^{o}pital's rule and this equation becomes equivalent to the condition $A \leq 0$, as expected. A similar calculation letting $x$ decrease to $\beta$ implies that $B \leq 0$. However, since $\brho$ is a density, we are looking for positive solutions. Hence, either $A=0$ or $\alpha$ coincides with the left endpoint of $\Om$. Similarly, either $B=0$ or $\beta$ coincides with the right edge of $\Om$. This is consistent with the result (\ref{eq:nodeltas2}) which showed that $\delta$-functions cannot occur in the interior of $\Om$.

In summary, we come to two conclusions:
\begin{itemize}
\item{A globally minimizing solution $\brho$ contains a $\delta$-function only if a boundary of the support of the solution coincides with a boundary of the domain.}
\item{A globally minimizing solution $\brho$ must satisfy (\ref{eq:mincon}).}
\end{itemize} 

We now consider three concrete examples for $\Om$ and $F(x)$.

\subsection{Example: Bounded interval}
\label{sec:bounded}

We model a one-dimensional biological swarm with repulsive social interactions described by the Laplace potential. We begin with the simplest possible case, namely no exogenous potential, $F(x)=0$ and a finite domain which for convenience we take to be the symmetric interval $\Om=[-d,d]$. As $F^{\prime \prime}(x) = 0$ we know from (\ref{eq:singlecomponent}) that the minimizing solution has a connected support, \emph{i.e.}, it is a single component. We will see that the minimizing solution has an equipartition of mass between $\delta$-functions at the boundaries of the domain and a constant solution in the interior, as shown schematically in Figure \ref{fig:repulsion_schematic}(a).

We now proceed with calculating the solution. From (\ref{eq:FIE-sol}), we find that
\begin{equation}
\brho_*(x) = \frac{\lambda}{2}.
\end{equation}
The nonclassical solution is
\begin{equation}
\brho(x) = \brho_* + A \delta(x+d) +B\delta(x-d) \quad \textrm{in $\Op$.}
\end{equation}
Eq. (\ref{eq:Mform}) gives $\lambda$ as
\begin{equation}
\lambda = \frac{M}{1+d}.
\end{equation}
Eq. (\ref{eq:AB}) specifies $A$ and $B$ as
\begin{equation}
A= B = \frac{M}{2(1+d)}.
\end{equation}
Since $\Om=\Op$, it follows that $\Lambda(x)=\lambda$ and $W_1$ vanishes according to (\ref{eq:W1Lambda}). Therefore, the solution is a global minimizer.

\begin{figure}
\centerline{\resizebox{!}{2.25in}{\includegraphics{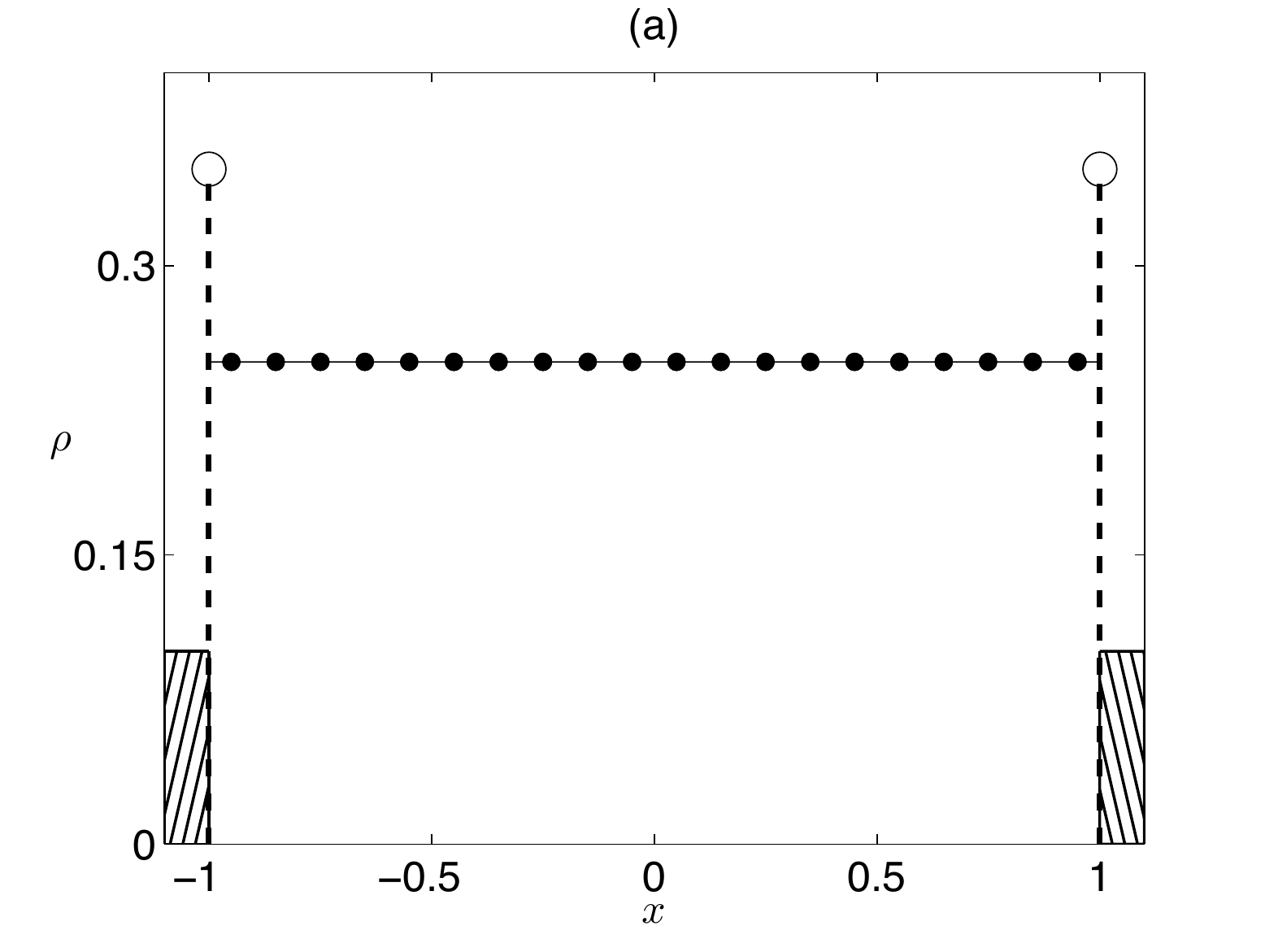}}}
\centerline{\resizebox{!}{2.25in}{\includegraphics{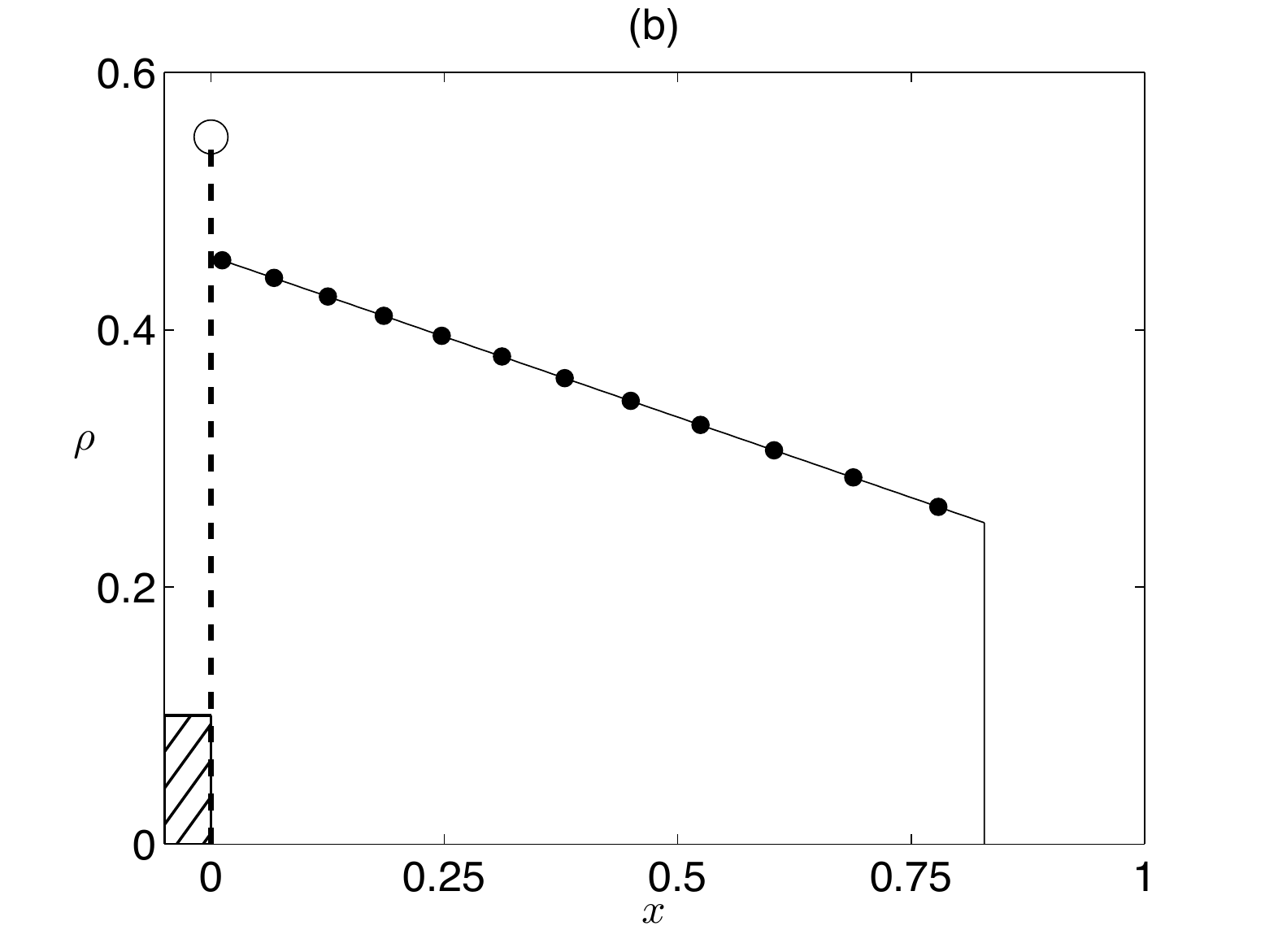}}}
\centerline{\resizebox{!}{2.25in}{\includegraphics{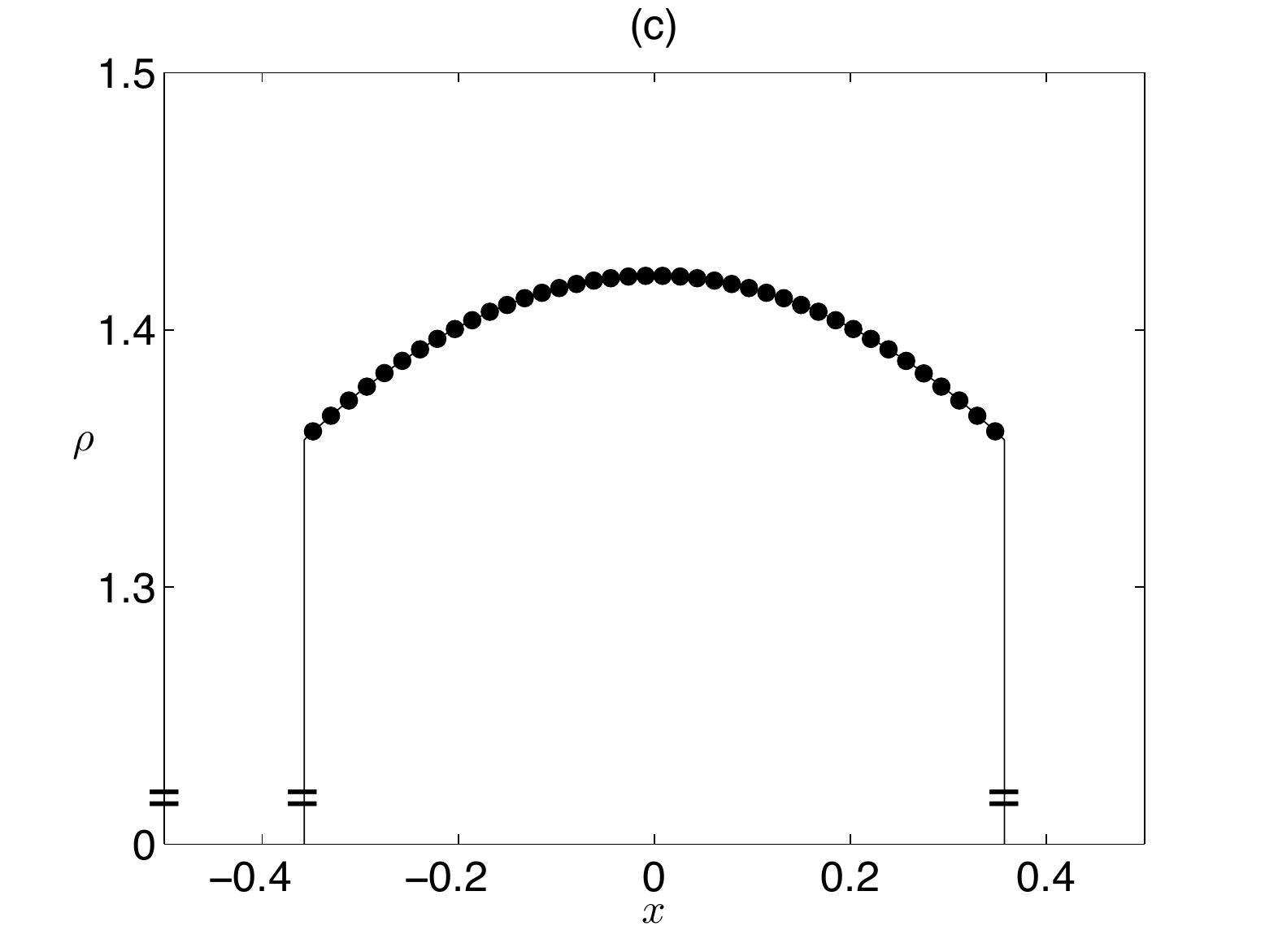}}}
\caption{Comparison of numerical and exact solutions of the minimization problem (\ref{eq:FIE}) and (\ref{eq:mass1}) when the potential $Q(x)$ is the repulsive Laplace potential~(\ref{eq:laplace}). The total mass is $M=1$. Dots correspond to equilibria of numerical simulations of the underlying discrete system (\ref{eq:discretesystem}) with $N=40$ swarm members. The density at the location of each swarm member is estimated using the correspondence of Section \ref{sec:contmodel}, visualized in Figure~\ref{fig:delta_schematic}. The solid curves represent exact solutions.~(a)~Bounded domain $\Om = [-1,1]$ with no exogeneous potential, \emph{i.e.}, $F(x) = 0$. Cross-hatched boxes indicate the domain boundaries. The ``lollipops'' at each end are  drawn at an arbitrary height and represent $\delta$-functions each of mass $10/40$ in the exact continuum solution, and a superposition of $10$ swarm members each in the simulation. See Section \ref{sec:bounded} and Figure \ref{fig:repulsion_schematic}(a).~(b)~The half-line $\Om = [0,\infty]$ with an exogenous gravitational potential $F(x) = gx$, $g=0.5$. The $\delta$-function at the origin has mass $28/40$. See Section \ref{sec:grav} and Figure \ref{fig:repulsion_schematic}(e,f).~(c) Unbounded domain $\Om = \mathbb{R}$ with an exogenous quadratic potential $F(x)=\gamma x^2$, $\gamma = 1$. The compactly supported solution drops discontinuously to zero, and the vertical axis is broken for convenience of visual display. See Section \ref{sec:quad} and Figure \ref{fig:repulsion_schematic}(bc).\label{fig:repulsion_numerics}}
\end{figure}

The solution $\brho(x)$ is shown schematically in Figure \ref{fig:repulsion_schematic}(a). Figure \ref{fig:repulsion_numerics}(a) compares analytical and numerical results for an example case where we take the total mass to be $M = 1$ and the finite domain to be $\Om = [-d,d]$ with $d=1$. Cross-hatched boxes indicate the boundary of the domain. The solid line is the classical solution $\brho$. Dots correspond to the numerically-obtained equilibrium of the discrete system (\ref{eq:discretesystem}) with $N=40$ swarm members. The density at each Lagrangian grid point is estimated using the correspondence discussed in Section (\ref{sec:contmodel}) and pictured in Figure~\ref{fig:delta_schematic}. Each ``lollipop'' at the domain boundary corresponds to a $\delta$-function of mass $10/N \cdot M = 1/4$ in the analytical solution, and simultaneously to a superposition of $10$ swarm members in the numerical simulation. Hence, we see excellent agreement between the continuum minimizer and the numerical equilibrium even for this relatively small number $N=40$ of Lagrangian points.

\subsection{Example: Gravitational potential on the half-line}
\label{sec:grav}

We  now consider repulsive social interactions and an exogenous gravitational potential. The spatial coordinate $x \geq 0$ describes the elevation above ground. Consequently, $\Om$ is the semi-infinite interval $0 \leq x < \infty$. Then $F(x)= gx$ with $g>0$, shown in Figure \ref{fig:repulsion_schematic}(f). As $F^{\prime \prime}(x) = 0$ we know from (\ref{eq:singlecomponent}) that the minimizing solution has a connected support, \emph{i.e.}, it is a single component. Moreover, translating this component downward decreases the exogenous energy while leaving the endogenous energy unchanged. Thus, the support of the solution must be $\Op = [0,\beta]$, potentially with $\beta = 0$. In fact, we will see that there are two possible solution types depending on $g$. For strong enough values of $g$, $\beta$ indeed equals zero, and the mass accumulates on the ground $x=0$, as shown schematically in Figure \ref{fig:repulsion_schematic}(d). For weaker $g$, the mass is partitioned between a $\delta$-function on the ground and a classical solution for $x>0$, as shown in Figure \ref{fig:repulsion_schematic}(e).

We now proceed with calculating the solution. From (\ref{eq:FIE-sol}), we find that
\begin{equation}
\brho_*(x) = \frac{\lambda-g x}{2}.
\end{equation}
The nonclassical solution is
\begin{equation}
\brho(x) = \brho_* + A \delta(x) +B\delta(x-\beta) \quad \textrm{in $\Op$.}
\end{equation}
Eq. (\ref{eq:Mform}) gives $\lambda$ as
\begin{equation}
\label{eq:gravitylambda}
\lambda = \frac{2M}{2+\beta} + \frac{g\beta}{2}.
\end{equation}
Eq. (\ref{eq:AB}) specifies $A$ and $B$ as
\begin{equation}
A= \frac{M}{2+\beta} + \frac{g}{2}, \quad B= \frac{M}{2+\beta} - \frac{g}{4}(2+\beta).
\end{equation}
For $\brho$ to be a global minimizer, it must be true that $B=0$ as shown in Section \ref{sec:funcmin}. Solving, we find that $\beta = 2 \sqrt{M/g} - 2$.

For $M>g$, we see that $\beta > 0$ in which case the minimizing solution is
\begin{equation}
\brho(x) =
\begin{cases}
\sqrt{gM} - \frac{g}{2}\left(1+x\right) + \sqrt{gM}\delta(x) &  0 \leq x \leq 2\sqrt{M/g} - 2 \\
0 & x >  2\sqrt{M/g} - 2.
\end{cases}
\end{equation}
It follows directly that $\brho(x) \geq 0$. Figure \ref{fig:repulsion_schematic}(e) shows a schematic drawing of this solution.

We still must consider the condition (\ref{eq:minconb}). To see it is satisfied, first note that $\lambda = g(\beta+1)$. Then (\ref{eq:minconb}) becomes
\begin{equation}
\quad \frac{F(x)e^x - F(\beta)e^\beta}{e^x-e^\beta} \geq g(1+\beta).
\end{equation}
We note that by l'H\^{o}pital's Rule that equality is obtained as $x \to \beta$. To show the inequality holds for $x > \beta$ let $x = \ln s$ and $\beta = \ln r$ (so $s > r > 0$). The inequality becomes
\begin{equation}
\quad \frac{sF(\ln s) - r F(\ln r)}{s - r} \geq g(1+\beta) \quad \textrm{for $s > r$}.
\end{equation}
We can interpret the left-hand side as the slope of a chord connecting the points $(r,r F(\ln r))$ and $(s, s F(\ln s))$. Consequently, if the function $sF(\ln s)$ is concave upward, the slope of the chord will be increasing as $s$ increases away from $r$, and the inequality will hold. Recalling that $F(x) = gx$, we compute $\left[sF(\ln s) \right]^{\prime \prime} = g/s$ which is positive, and hence the solution is globally stable.

For $M \leq g$ our previous calculation naively implies $\beta \leq 0$. Since $\beta$ cannot be negative, the minimizer in this case is a $\delta$-function at the origin, namely $\brho(x) = M\delta(x)$, shown in Figure \ref{fig:repulsion_schematic}(d). In this case, $\lambda = M$ from (\ref{eq:gravitylambda}) and $\Lambda(x) = M \exp{-x} + gx$ from (\ref{eq:lambdaright}). It follows that
\begin{equation}
\Lambda(x) = M \exp{-x} + gx \geq M(1-x) + gx = M + (g - M) x \geq M \quad \textrm{for $x \geq 0$}.
\end{equation}
The first inequality follows from a Taylor expansion. The second follows from our assumption $ M \leq g$. Since $\Lambda(x) \geq \lambda$ the solution is a global minimizer.

In summary, there are two cases.  When $M \leq g$, the globally stable minimizer is a $\delta$-function at the origin. When $M > g$ there is a globally stable minimizer consisting of a $\delta$-function at the origin which is the left-hand endpoint of a compactly-supported classical swarm. The two cases are shown schematically in Figures \ref{fig:repulsion_schematic}(de). Figure \ref{fig:repulsion_numerics}(b) compares analytical and numerical results for the latter ($M > g$) case with  $M = 1$ and $g = 0.5$. We use $N=40$ swarm members for the numerical simulation. The numerical (dots) and analytical (line) $\bsrho$ agree, as does the nonclassical part of the solution, pictured as the ``lollipop'' which represents a superposition of $28$ swarm members in the numerical simulation having total mass $28/N \cdot M = 0.7$, and simultaneously a $\delta$-function of mass $\sqrt{gM}=1/\sqrt{2}\approx 0.707$ in the analytical solution.

\subsection{Example: Quadratic potential on the infinite line}
\label{sec:quad}

We now consider the infinite domain $\Om = \mathbb{R}$ with a quadratic exogenous potential well, pictured in Figure \ref{fig:repulsion_schematic}(c). This choice of a quadratic well is representative of a generic potential minimum, as might occur due to a chemoattractant, food source, or light source. Thus $F(x)= \gamma x^2$ where $\gamma>0$ controls the strength of the potential. As $F^{\prime \prime}(x) = 2\gamma > 0$ we know from (\ref{eq:singlecomponent}) that the minimizing solution is a single component. We take the support of the solution to be $\Op = [\alpha,\beta]$. We will find that the compactly supported density is classical and has an inverted parabolic shape, shown in Figure \ref{fig:repulsion_schematic}(b).

Our calculation proceeds as follows. We know from Section \ref{sec:funcmin} that because $\alpha$ and $\beta$ are assumed to be finite, $A=B=0$ and so the solution is classical. From~(\ref{eq:FIE-sol}), we find that
\begin{equation}
\brho = \brho_*(x) = \frac{\lambda + 2 \gamma -\gamma x^2}{2},
\end{equation}
and, from (\ref{eq:AB}),
\begin{equation}
A= \frac{\lambda - \gamma \alpha^2 + 2\gamma \alpha}{2} = 0, \quad
B= \frac{\lambda - \gamma \beta^2 - 2\gamma \beta}{2} = 0. 
\label{eq:ABquad}
\end{equation}
Eliminating $\lambda$ from these equations and recalling that $\beta \geq \alpha$, It follows that $\alpha = -\beta$. Hence, the solution is symmetric around the center of the potential. For convenience, we now define $\beta = -\alpha = H$.

Eq. (\ref{eq:Mform}) gives $\lambda$ in terms of the mass $M$ as
\begin{equation}
\lambda = \frac{M + \gamma \left(H^2 + H^3/3 \right)}{1+H}.
\end{equation}
However, from the second half of (\ref{eq:ABquad}) we know that
\begin{equation}
\lambda = \gamma(H+1)^2 - \gamma.
\end{equation}
Equating these two expressions for $\lambda$ yields that
\begin{equation}
H = \left(\frac{3M}{2\gamma} + 1\right)^{1/3} - 1.
\end{equation}
Note that $H$ increases monotonically from $0$ with increasing $M$.

Finally, writing the solution $\brho$ in terms of $M$, we have
\begin{equation}
\brho(x) =
\begin{cases}
\frac{\gamma}{2} \left[ \left( \frac{3M}{2\gamma}+1\right)^{2/3}+ \left(1-x^2\right)\right] &  |x| \leq  \left(\frac{3M}{2\gamma} + 1\right)^{1/3} - 1 \\
0 & |x| >   \left(\frac{3M}{2\gamma} + 1\right)^{1/3} - 1.
\end{cases}
\end{equation}
It follows directly that $\brho(x) \geq 0$, and that there is a discontinuity at the edge of the support.

We now must show that the candidate $\brho$ is a global minimizer, which is done by demonstrating (\ref{eq:mincon}). By the same argumentation used for the right endpoint in Section \ref{sec:grav}, it suffices to show that $sF(\ln s)$ is concave up for $s > \exp{H}$ which follows directly from algebra. Hence, the solution is a global minimizer.

The solution $\brho$ is shown schematically in Figure \ref{fig:repulsion_schematic}(b). Figure \ref{fig:repulsion_numerics}(b) compares analytical and numerical results for the case with  $M = 1$ and $\gamma = 1$, with $N = 40$ swarm members used for the numerical solution. The numerical (dots) and analytical (line) $\brho$ agree closely.

\section{Examples with a Morse-type social force}
\label{sec:morse}


In many physical, chemical, and biological applications, the pairwise potential $Q$ is isotropic, with  a repulsion dominating at short distance and the interaction strength approaching zero for very long distances. A common choice for $Q$ is the Morse potential 
\begin{equation}
Q(x) =  -\F L\exp{-|x|/L} +\exp{-|x|}, \label{eq:Morse}
\end{equation}
with appropriately chosen parameters. Here, $x$ is the distance between particles, $L>1$ is the characteristic length scale of attraction, and $\F<1$ is the characteristic velocity induced by attraction. We have scaled the characteristic repulsive strength and length scale to be unity. In this section we are concerned with the solution of (\ref{eq:mass1}) and (\ref{eq:FIE}) with $Q$ given by (\ref{eq:Morse}).

The Morse potential has been studied extensively and has become a canonical model for attractive-repulsive interactions \cite{LevRapCoh2001,MogEdeBen2003,DOrChuBer2006,TopBerLog2008}. A key characteristic of the potential (\ref{eq:Morse}) is whether or not the parameters $\F,L$ are chosen to be in the \emph{H-stable} or the \emph{catastrophic} regime; see \cite{Rue1969} for a review. Consider (\ref{eq:discretesystem}) with $F(x)=0$. If the parameters $\F,L$ are chosen in the H-stable regime $\F L^2 < 1$ , then as the number of particles $N$ increases, the density distribution of particles approaches a constant, as does the energy per particle. Stated differently, the particles form a crystalline lattice where the nearest-neighbor distance is approximately equal for all particles (excluding edge effects). As more individuals are added to the group, the inter-organism spacing is preserved and the group grows to cover a larger spatial region. If the parameters are chosen outside of the H-stable regime, \emph{i.e.}, $\F L ^2 > 1$, the system is catastrophic. In this case, the energy per particle is unbounded as $N \rightarrow \infty$, and particles pack together more and more closely as N increases.

Our work in \cite{LevTopBer2009} classifies the asymptotic behaviors of the continuum problem~(\ref{eq:pde}) with Morse-type interactions in the absence of an external potential, $F(x) = 0$, and on the real line. In the H-stable regime, the continuum model displays spreading density profiles, while in the catastrophic regime, it forms compactly-supported steady states.

The distinction between catastrophic and H-stable is related to the Fourier transform of the potential $Q(x)$. Note that (\ref{eq:Morse}) has Fourier transform
\begin{subequations}
\label{eq:qhatmorse}
\begin{eqnarray}
\widehat{Q}(k) &=&  
\frac{2}{ 1+{k}^{2} }-\frac {2\F L^2}{1+(kL)^2}\\
&=& \frac{2[(1-\F L^2)+L^2(1-\F)k^2]}{(1+{k}^{2})(1+(kL)^2)}.
\end{eqnarray}
\end{subequations}
The condition for H-stability is equivalent to $\widehat{Q}(0)  = 2(1- \F L^2) > 0$. In this case, $\widehat{Q}(k) > 0$ for all $k$, and from Section \ref{sec:minimizers}, this is sufficient for $W_2 > 0$, or equivalently, linear stability of constant density states. In the catastrophic case, $\widehat{Q}(0) < 0$. Intuitively, in the catastrophic case, the constant density state is unstable to long wave perturbations. The system is attracted to states of finite extent shorter than the length scale of the instability. In the H-stable case, the constant density state is stable to perturbations, and initial density profiles spread evenly to become flat. We will now study minimizers for the case of Morse-type interactions, and will see qualitatively different solutions for the catastrophic and H-stable cases.

\subsection{Classical and nonclassical solutions}
\label{sec:morsenonclass}

We follow the procedure used in Section \ref{sec:repulsive}, namely we first look for a classical solution on the interior of $\Op=[\alpha,\beta]$ and then allow for $\delta$-functions at the boundaries. Once again, we will see that minimizers contain $\delta$-functions only when the boundary of the support, $\Op$, coincides with the boundary of the domain, $\Om$.

For convenience, define the differential operators  $\mathcal{L}_1 \equiv  \partial_{xx} - 1 $ and 
$\mathcal{L}_2 \equiv  L^2\partial_{xx} - 1$, and apply $-\mathcal{L}_1 \mathcal{L}_2$ to (\ref{eq:FIE}) to obtain
\begin{equation}
\nu (\bsrho)_{xx} + \epsilon \bsrho = -\mathcal{L}_1 \mathcal{L}_2 [\lambda - F(x)], \quad x \in \Op, \label{eq:morselocal}
\end{equation}
where
\begin{equation}
\nu = 2L^2(1-\F) > 0, \quad \epsilon = 2(\F L^2-1).
\end{equation}
Thus, we guess
\begin{equation}
\label{eq:morseansatz}
\rho(x) = \bsrho(x) + A \delta(x-\alpha) + B \delta(x-\beta).
\end{equation}
The full solution to the problem is obtained by substituting (\ref{eq:morseansatz}) into (\ref{eq:mass1}) and (\ref{eq:FIE})  which yields 
\begin{equation}
\label{eq:morseintegraleq}
\int_\alpha^\beta Q(x-y)  \bsrho(y)~dy + AQ(x-\alpha) + BQ(x-\beta) = \Lambda(x) - F(x),
\end{equation}
where $\Lambda(x) = \lambda$ in $\Op$. We begin by considering the amplitudes $A$ and $B$ of the distributional component of the solution. We factor the differential operators  $\mathcal{L}_1 \equiv  \partial_{xx} - 1 = \cPp \cPm = \cPm \cPp $ 
where $ \cPpm = \partial_x \pm 1$ and $\mathcal{L}_2 \equiv  L^2\partial_{xx} - 1 = \cQp \cQm = \cQm \cQp $
where $ \cQpm = L \partial_x \pm 1$. Note that
\begin{equation}
\cPm \cQm Q(x-y)=2L(\F-1) \delta(x-y) -2(L+1)\left[\F\exp{(y-x)/L}-\exp{y-x}\right ] H(x-y),
\end{equation}
where $H$ is the Heaviside function. 
Now we apply $\cPm \cQm $ to (\ref{eq:morseintegraleq}) at a point $x$ in $\Op$, which yields
\begin{eqnarray}
2L(\F-1)\bsrho(x)
- 2(L+1) \int_\alpha^x \left [(\F\exp{(y-x)/L}-\exp{y-x} \right ]  \bsrho(y)\,dy && \\
  \mbox{}+A \cPm \cQm Q(x-\alpha) = \cPm \cQm \{ \lambda -F(x)\}. \nonumber
\end{eqnarray}
Taking the limit $x \to \alpha^+$ yields
 \begin{equation}
\label{eq:morseboundary1}
 2L(\F-1)\bsrho(\alpha)+2A (L+1)(1-\F) = \lambda -  \cPm \cQm F(x) \Big |_{x=\alpha},
 \end{equation}
where we have used the fact that $ \cPm \cQm \lambda = \lambda$. A similar calculation using the operators $\cPp \cQp$ and focusing near $x=\beta$ yields that
\begin{equation}
\label{eq:morseboundary2}
2L(\F-1)\bsrho(\beta)+2B (L+1)(1-\F) = \lambda -  \cPp \cQp F(x) \Big |_{x=\beta}.
\end{equation}
Eqs. (\ref{eq:morseboundary1}) and (\ref{eq:morseboundary2}) relate the amplitudes of the $\delta$-functions at the boundaries to the value of the classical solution $\bsrho$ there. Further solution of the problem requires $F(x)$ to be specified. 

In the case where $\Om = \Op$, solving (\ref{eq:morselocal}) for $\bsrho(x)$ and solving (\ref{eq:morseboundary1}) and (\ref{eq:morseboundary2}) for $A$ and $B$ yields an equilibrium solution. One must check that the solution is non-negative and then consider the solutions stability to determine if it is a local or global minimizer. In the case where $\Op$ is contained in the interior of $\Om$, we know that $A = B = 0$ as discussed in Section \ref{sec:absence}. We consider this case below.

Suppose $\Op$ is contained in the interior of $\Om$. Then $A = B = 0$. Following Section~\ref{sec:funcmin}, we try to determine when $\Lambda(x) \geq \lambda$ in $\Omega$ and when $W_2 > 0$, which constitute necessary and sufficient conditions for $\brho$ to be a global minimizer. We apply $\cPm \cQm$ to (\ref{eq:morseintegraleq}) at a point $x<\alpha$. The integral term and the terms arising from the $\delta$ functions vanish. The equation is simply $\cPm \cQm \{ \Lambda(x) - F(x) \} = 0$. We write the solution as
\begin{equation}
\label{eq:lambdamorse}
\Lambda(x) = F(x) + k_1 \exp{x-\alpha} + k_2 \exp{(x-\alpha)/L}.
\end{equation}
The two constants $k_{1,2}$ are determined as follows. From (\ref{eq:morseintegraleq}), $\Lambda(x)$ is a continuous function, and thus
\begin{equation}
\label{eq:lambdacontinuity}
\Lambda(\alpha) = F(\alpha) + k_1 + k_2 = \lambda.
\end{equation}
We derive a jump condition on the derivative to get another equation for $k_{1,2}$. We differentiate (\ref{eq:morseintegraleq}) and determine that $\Lambda^\prime(x)$ is continuous. However, since $\Lambda(x) = \lambda$ for $x \in \Op$, $\Lambda^\prime(\alpha) = 0$. Substituting this result into the derivative of (\ref{eq:lambdamorse}) and letting $x$ increase to $\alpha^-$, we find
\begin{equation}
\label{eq:lambdaprime}
\Lambda^\prime(\alpha^-) = F^\prime(\alpha) + k_1 + k_2/L = 0.
\end{equation}
The solution to (\ref{eq:lambdacontinuity}) and (\ref{eq:lambdaprime}) is
\begin{subequations}
\label{eq:kvals}
\begin{eqnarray}
k_1 & = & \frac{1}{L-1} \bigl\{   F(\alpha) - LF^\prime(\alpha) - \lambda \bigr\}, \\
k_2 & = & \frac{L}{L-1} \bigl\{   -F(\alpha) + F^\prime(\alpha) + \lambda  \bigr\}.
\end{eqnarray}
\end{subequations}

Now that $\Lambda(x)$ is known near $x=\alpha$ we can compute when $\Lambda(x) \geq \lambda$, at least near the left side of $\Op$. Taylor expanding $\Lambda(x)$ around $x=\alpha$, we find
\begin{equation}
\label{eq:lambdatayl}
\Lambda(x) \approx \lambda + \frac{1}{2L} \Bigl\{ \cPm \cQm F(x) \big |_{x=\alpha} - \lambda  \Bigr \} (x-\alpha)^2 + \ldots .
\end{equation}
The quadratic term in (\ref{eq:lambdatayl}) has coefficient
\begin{subequations}
\begin{eqnarray}
& &  \frac{\cPm \cQm F(x) \big |_{x=\alpha} - \lambda}{2L}, \\
& = & (1-\F) \bsrho(\alpha) \geq 0,
\end{eqnarray}
\end{subequations}
where the second line comes from substituting (\ref{eq:morseboundary1}) with $A=0$ and noting that the classical part of the solution $\bsrho(\alpha)$ must be nonnegative since it is a density. Furthermore, since we expect $\bsrho(\alpha) > 0$ (this can be shown a posteriori), we have that the quadratic term in (\ref{eq:lambdatayl}) is positive.

A similar analysis holds near the boundary $x=\beta$. Therefore,  $\Lambda(x) \ge \lambda$ for $x$ in a neighborhood outside of $\Op$. Stated differently, the solution (\ref{eq:morseansatz}) is a swarm minimizer, that is, it is stable with respect to infinitesimal redistributions of mass. The domain $\Op$ is determined through the relations (\ref{eq:morseboundary1}) and (\ref{eq:morseboundary2}), which, when $A=B=0$, become
\begin{subequations}
\label{eq:bcs}
\begin{eqnarray}
2L(\F-1)\bsrho(\alpha) & = & \lambda -  \cPm \cQm F(x) \Big |_{x=\alpha}, \\
2L(\F-1)\bsrho(\beta)   & = & \lambda -  \cPp  \cQp  F(x) \Big |_{x=\beta}.
\end{eqnarray}
\end{subequations}

In the following subsections, we will consider the solution of the continuum system (\ref{eq:FIE}) and (\ref{eq:mass1}) with no external potential, $F(x) = 0$. We consider two cases for the Morse interaction potential (\ref{eq:Morse}): first, the catastrophic case on $\Om = \mathbb{R}$, for which the above calculation applies, and second, for the H-stable case on a finite domain, in which case $A,B \neq 0$ and there are $\delta$-concentrations at the boundary. Exact solutions for cases with an exogenous potential, $F(x) \neq 0 $ can be straightforwardly derived, though the algebra is even more cumbersome and the results unenlightening.

\begin{figure}
\centerline{\resizebox{!}{2.5in}{\includegraphics{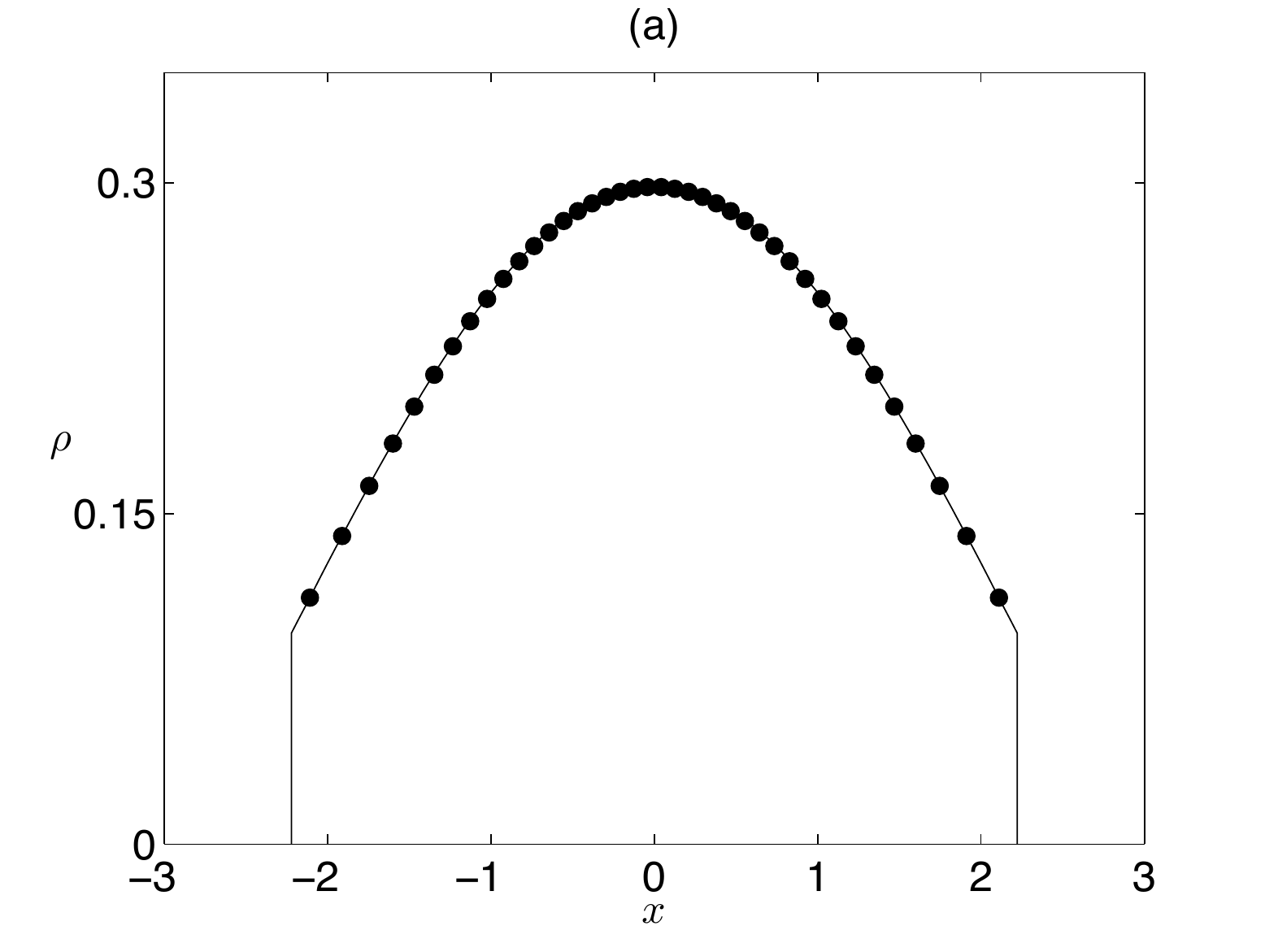}}}
\centerline{\resizebox{!}{2.5in}{\includegraphics{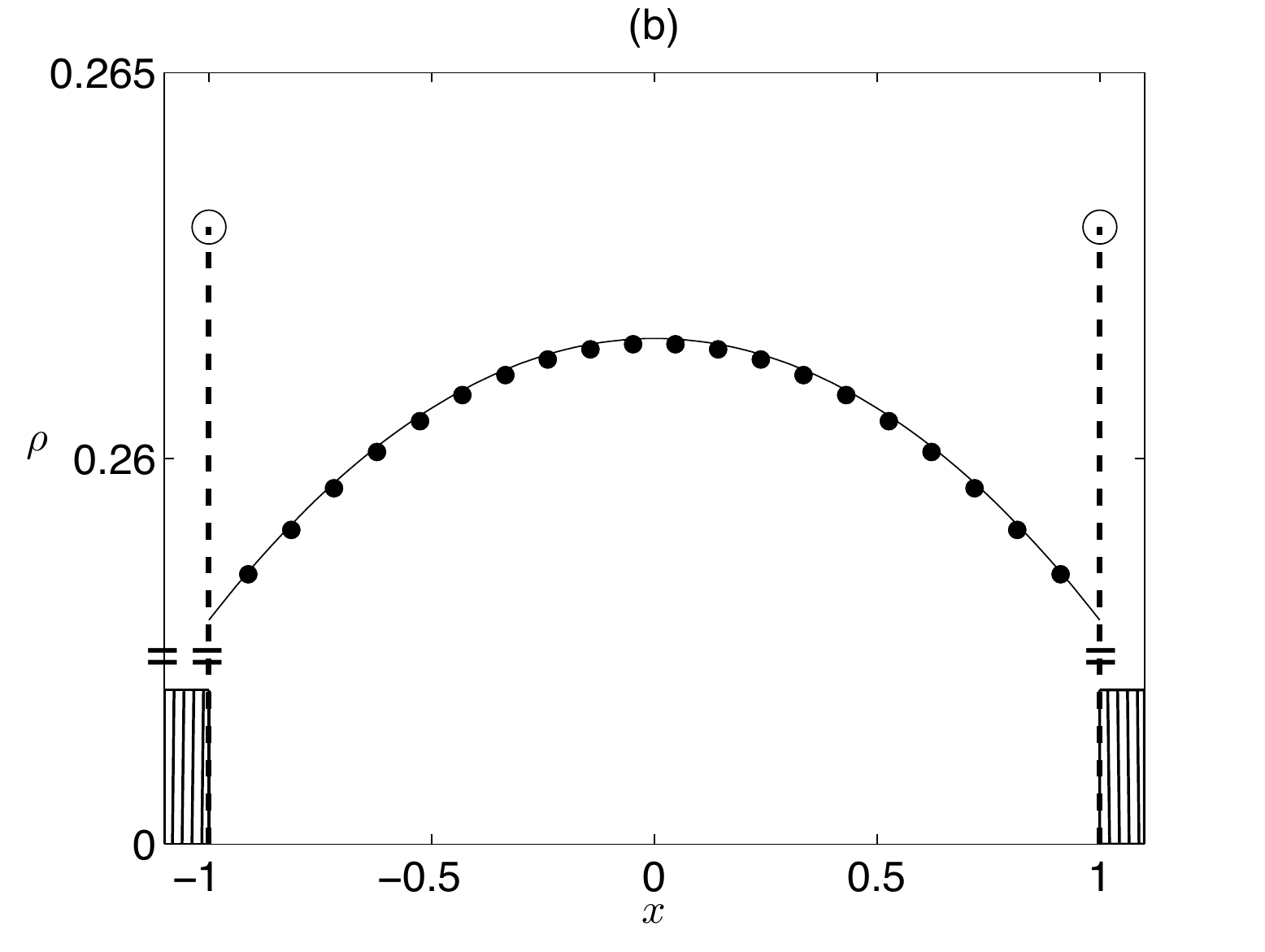}}}
\caption{Comparison of numerical and exact solutions of the minimization problem (\ref{eq:FIE}) and (\ref{eq:mass1}) when the potential $Q(x)$ is the Morse potential~(\ref{eq:Morse}). The total mass is $M=1$. Dots correspond to equilibria of numerical simulations of the underlying discrete system (\ref{eq:discretesystem}) with $N=40$ swarm members. The solid curves represent exact solutions.~(a)~Unbounded domain $\Om = \mathbb{R}$ with parameters $F=0.5$, $L=2$ in (\ref{eq:Morse}), which is in the catastrophic parameter regime. See Section \ref{sec:catastrophic}.~(b) Bounded domain $\Om = [-1,1]$ with $F=0.1$, $L=2$, which is in the H-stable parameter regime. The $\delta$-functions at the boundary each have mass $10/40$ and the vertical axis is broken for convenience of visual display. See Section \ref{sec:Hstable}. \label{fig:morse_numerics}}
\end{figure}

\subsection{Example: Catastrophic interactions in free space}
\label{sec:catastrophic}

In this case, $F(x) = 0$ in (\ref{eq:FIE}) and $\F L^2 > 1$ in (\ref{eq:Morse}) so that $\epsilon > 0$ in (\ref{eq:morselocal}). The solution to (\ref{eq:morselocal}) is
\begin{equation}
\label{eq:freespacerhobar}
\brho(x) = \bsrho(x) \equiv C \cos(\mu x) + D \sin(\mu x) - \lambda / \epsilon,
\end{equation}
where
\begin{equation}
\mu = \sqrt{\frac{\epsilon}{\nu}} = \sqrt{\frac{\F L^2 - 1}{L^2(1-\F)}}.
\end{equation}
In the absence of an external potential, the solution is translationally invariant. Consequently, we may choose the support to be an interval $\Op = [-H,H]$ which is symmetric around the origin. Hence, by symmetry, $D=0$. While the solution (\ref{eq:freespacerhobar}) satisfies the ordinary differential equation (\ref{eq:morselocal}), substituting into the integral equation (\ref{eq:FIE}) and the mass constraint (\ref{eq:mass1}) will determine the constants $C$, $\lambda$ and $H$. The integral operator produces modes spanned by $\{ \cosh{z}, \cosh{z/L}\}$. From these modes follow two homogeneous equations for $C$ and $\lambda$ which simplify to
\begin{subequations}
\begin{eqnarray}
\frac{2}{1+\mu^2}\left[ \cos( \mu H) - \mu \sin (\mu H) \right]C - \frac{1}{\F L^2 -1}\lambda & = & 0, \label{eq:catastrophiceig1} \\
\frac{2}{1+\mu^2 L^2}\left[ \cos( \mu H) - \mu L \sin (\mu H) \right]C - \frac{1}{\F L^2 -1}\lambda & = & 0.
\end{eqnarray}
\end{subequations}
For these equations to have a nontrivial solution for $C$ and $\lambda$ the determinant of the coefficient matrix must vanish, which yields a condition specifying $H$,
\begin{equation}
\cot(\mu H) =  \frac{\F L-1}{\sqrt{(1-\F)(\F L^2-1)}}.
\end{equation}
The mass constraint (\ref{eq:mass1}) yields
\begin{equation}
\label{eq:catastrophicmass}
\frac{2}{\mu}\sin(\mu H) C - \frac{H}{\F L^2 - 1}\lambda = M.
\end{equation}
Solving (\ref{eq:catastrophiceig1}) and (\ref{eq:catastrophicmass}) for $C$ and $\lambda$ yields the full solution for the coefficients in (\ref{eq:freespacerhobar}) and the half-width $H$ of the solution,
\begin{subequations}
\begin{eqnarray}
H & = & \frac{1}{\mu}\cot^{-1}_{[0,\pi]} \left\{ \frac{\F L-1}{\sqrt{(1-\F)(\F L^2-1)}} \right\}, \\
C & = & \frac{M}{2(H+L+1)} \frac{\sqrt{\F}(L^2-1)}{L(1-\F)}, \\
\lambda &= &\frac{M(1-\F L^2)}{H + L+1}.
\end{eqnarray}
\end{subequations}

As we have shown in Section \ref{sec:morsenonclass}, this solution is a swarm minimizer. In fact, the solution is also a local minimizer. To see this, note that from (\ref{eq:lambdamorse}) and (\ref{eq:kvals}) that
\begin{equation}
\Lambda(x) = 
\begin{cases}
\lambda & |x| < H \\
\frac{\lambda}{L-1} \left[ L \exp{(|x|-H)/L} - \exp{|x|-H}\right] & |x| \geq H.
\end{cases}
\end{equation}
Since $\lambda < 0$, we see that $\Lambda(x) > \lambda$ for $|x| > H$, ensuring that the solution is a local minimizer.

While we suspect that $\bsrho$ is a global minimizer, this is not immediately apparent because $\widehat{Q}(k)$  in (\ref{eq:qhatmorse}) has mixed sign in this catastrophic case, and hence $W_2$ is of indeterminate sign. To establish that $\bsrho$ is a global minimizer one might study the quantity $W_1 + W_2$ but we leave this analysis as an open problem.

Figure \ref{fig:morse_numerics}(a) compares analytical and numerical results for an example case with total mass $M = 1$ and interaction potential parameters $F = 0.5$ and $L=2$. The solid line is the compactly supported analytical solution $\brho$. Dots correspond to the numerically-obtained equilibrium of the discrete system (\ref{eq:discretesystem}) with $N=40$ swarm members.

\subsection{Example: H-stable interactions on a bounded domain}
\label{sec:Hstable}

As described in \cite{LevTopBer2009}, the asymptotic behavior of (\ref{eq:pde}) for the H-stable case is a spreading self-similar solution that approaches the well-known Barenblatt solution of the porous medium equation. Hence, there is no equilibrium solution for the H-stable case on an unbounded domain (one can verify this by considering the analogous problem to that of the previous section and showing explicitly that there is no solution). Here, we assume a bounded domain $\Om = [-d,d]$. As before, $F(x) = 0$ in (\ref{eq:FIE}) but now $\F L^2 < 1$ in (\ref{eq:Morse}) so that $\epsilon < 0$ in (\ref{eq:morselocal}). The classical solution to (\ref{eq:morselocal}) is
\begin{equation}
\label{eq:hstablerhobar}
\bsrho(x) \equiv C \cosh(\tmu x) + D \sinh(\tmu x) - \lambda / \epsilon,
\end{equation}
where
\begin{equation}
\tmu = \sqrt{\frac{-\epsilon}{\nu}} = \sqrt{\frac{1 - \F L^2}{L^2(1-\F)}}.
\end{equation}
We will again invoke symmetry to assume $D=0$. The minimizer will be the classical solution together with $\delta$-functions on the boundary,
\begin{equation}
\brho(x) = \bsrho(x) + A \delta (x+d) + B \delta (x-d).
\end{equation}
Again by symmetry, $B=A$. Consequently, the solution can be written as
\begin{equation}
\brho = C \cosh(\tmu x) - \lambda/\epsilon + A [\delta(x+d) + \delta(x-d)].
\end{equation}

Substituting into the integral equation (\ref{eq:FIE}) and the mass constraint (\ref{eq:mass1}) will determine the constants $C$, $\lambda$ and $A$. The integral operator produces modes spanned by $\{ \cosh{z}, \cosh{z/L}\}$. This produces two homogeneous, linear equations for $C$, $A$ and $\lambda$. The mass constraint (\ref{eq:mass1}) produces an inhomogeneous one, namely an equation linear in $C$, $A$, and $\lambda$ for the mass. We have the three dimensional linear system
\begin{equation}
\begin{bmatrix}
-1 & \displaystyle \frac{1}{2} \left\{ \displaystyle \frac{\exp{\tmu d}}{1-\tmu} + \displaystyle \frac{\exp{-\tmu d}}{1+\tmu} \right\} & \displaystyle \frac{1}{2(1 - \F L^2 )}\medskip \\
-1 & \displaystyle \frac{L}{2} \left\{ \displaystyle \frac{\exp{\tmu d}}{1-\tmu L} + \displaystyle \frac{\exp{-\tmu d}}{1+\tmu L} \right\} & \displaystyle \frac{L}{2(1 - \F L^2 )} \medskip \\
2 & \displaystyle \frac{2\sinh (\tmu d)}{\tmu} & \displaystyle \frac{d}{1 - \F L^2 }
\end{bmatrix}
\begin{bmatrix}
A  \\ C \\ \lambda
\end{bmatrix}
=
\begin{bmatrix}
0  \\ 0 \\ M
\end{bmatrix}.
\end{equation}
The solution is
\begin{subequations}
\label{eq:hstablesoln}
\begin{eqnarray}
A & = & \frac{\tmu ^2 L M \left[ (1+\tmu) (1+\tmu L) \exp{\tmu d} - (1-\tmu) (1-\tmu L) \exp{-\tmu d}\right]}{2\Phi}, \\
C & = & -\frac{\tmu M (1-\tmu^2) (1-\tmu^2 L^2)}{\Phi}, \\
\lambda & = & \frac{M \tmu (1-\F L^2)\left[ (1+\tmu)(1+\tmu L) + (1-\tmu)(1-\tmu L) \right]}{\Phi},
\end{eqnarray}
\end{subequations}
where for convenience we have defined
\begin{equation}
\Phi = (1+\tmu)(1+\tmu L) (\tmu L + \tmu d + \tmu - 1)\exp{\tmu d} - (1-\tmu)(1-\tmu L) (\tmu L + \tmu d + \tmu + 1)\exp{-\tmu d}.
\end{equation}

For this H-stable case, $\widehat{Q}(k) > 0$ which ensures that $W_2 > 0$ for nontrivial perturbations. This guarantees that the solution above is a global minimizer.

In the limit of large domain size $d$, the analytical solution simplifies substantially. To leading order, the expressions (\ref{eq:hstablesoln}) become
\begin{subequations}
\begin{eqnarray}
A & = & \frac{\tmu L M}{2 d}, \\
C & = & -\frac{\exp{-\tmu d}M(1-\tmu)(1-\tmu L)}{d},\\
\lambda & = & \frac{M(1-\F L^2)}{d}.
\end{eqnarray}
\end{subequations}
Note that $C \cosh(\tmu x)$ is exponentially small except in a boundary layer near each edge of $\Om$, and therefore the solution is nearly constant in the interior of $\Om$.

Figure \ref{fig:morse_numerics}(b) compares analytical and numerical results for an example case with a relatively small value of $d$. We take total mass $M = 1$ and set the domain half-width to be $d=1$. The interaction potential parameters $F = 0.5$ and $L=2$. The solid line is the classical solution $\bsrho$. Dots correspond to the numerically-obtained equilibrium of the discrete system (\ref{eq:discretesystem}) with $N=40$ swarm members. Each ``lollipop'' at the domain boundary corresponds to a $\delta$-function of mass $10/N \cdot M = 1/4$ in the analytical solution, and simultaneously to a superposition of $10$ swarm members in the numerical simulation.

\section{Modeling a locust swarm: Examples with a gravitational potential}
\label{sec:locust-ground}

We now return to the locust swarm model of \cite{TopBerLog2008}, discussed also in Section \ref{sec:intro}. Recall that locust swarms are observed to have a concentration of individuals on the ground, a gap or ``bubble'' where the density of individuals is near zero, and a sharply delineated swarm of flying individuals. This behavior is reproduced in the model (\ref{eq:locusts}); see Figure~\ref{fig:locust}(b). In fact, Figure~\ref{fig:locust}(c) shows that the bubble is present even when the wind in the model is turned off, and only endogenous interactions and gravity are present.

To better understand the structure of the swarm, we consider the analogous continuum problem. To further simplify the model, we note that the vertical structure of the swarm appears to depend only weakly on the horizontal direction, and thus we will construct a \emph{quasi-two-dimensional} model in which the horizontal structure is assumed uniform.

In particular, we will make a comparison between a one-dimensional and a quasi-two-dimensional model. Both models take the form of the energy minimization problem (\ref{eq:FIE}) on a semi-infinite domain, with an exogenous potential $F(x)=gx$ describing gravity. The models differ in the choice of the endogenous potential $Q$, which is chosen to describe either one-dimensional or quasi-two-dimensional repulsion. The one-dimensional model is precisely that which we considered in Section \ref{sec:grav}. There we saw that minimizers of the one-dimensional model can reproduce the concentrations of locusts on the ground and a group of individuals above the ground, but there cannot be a separation between the grounded and airborne groups. We will show below that for the quasi-two-dimensional model, this is not the case, and indeed, some minimizers have a gap between the two groups. 

As mentioned, the one-dimensional and quasi-two-dimensional models incorporate only endogenous repulsion. However, the behavior we describe herein does not change for the more biologically realistic situation when attraction is present. We consider the repulsion-only case in order to seek the minimal mechanism responsible for the appearance of the gap.



\subsection{The quasi-two-dimensional Laplace potential}

We consider a swarm in two dimensions, with spatial coordinate $\vec{x} =(x_1,x_2)$. We will eventually confine the vertical coordinate $x_1$ to be nonnegative, since it describes the elevation above the ground at $x_1 = 0$. We assume the swarm to be uniform in the horizontal direction $x_2$, so that $\rho(x_1,x_2) \equiv \rho(x_1)$.

We construct a quasi-two-dimensional interaction potential,
\begin{equation}
Q_{2D} (|x_1-y_1|) = - \int_{-\infty}^{\infty} Q(|\vec{x}-\vec{y}| ) ~ dy_2.
\end{equation}
Letting $z=x_1-y_1$ and $s = x_2 - y_2$, this yields
\begin{equation}
Q_{2D} (z) = \int_{-\infty}^{\infty} Q(\sqrt{s^2+z^2 }) ~ ds.
\end{equation}
It is straightforward to show that the two-dimensional energy per unit horizontal length is given by
\begin{equation}
\label{eq:W2d}
W_{2D}[\rho] = \frac{1}{2} \int_\Om \int_\Om \rho(x_1) \rho(y_1) Q_{2D}(x_1-y_1)\,dx_1\,dy_1 + \int_\Om F(x_1)\rho(x_1)\,dx_1,
\end{equation}
where the exogenous force is $F(x_1)=gx_1$ and the domain $\Om$ is the half-line $x_1 \geq 0$. This is exactly analogous to the one-dimensional problem (\ref{eq:continuum_energy}), but with particles interacting according to the quasi-two-dimensional endogenous potential. Similarly, the corresponding dynamical equations are simply (\ref{eq:cont_velocity}) and (\ref{eq:pde}) but with endogenous force $q_{2D} = -Q_{2D}^\prime$.

For the Laplace potential (\ref{eq:laplace}), the quasi-two-dimensional potential is
\begin{equation}
\label{eq:q2dlaplace}
Q_{2D} (z) =  \int_{-\infty}^{\infty} e^{-\sqrt{s^2+z^2 }} ~ ds.
\end{equation}
This integral can be manipulated for ease of calculation,
\begin{subequations}
\begin{eqnarray}
Q_{2D} (z) &=& -\int_{-\infty}^{\infty} e^{-\sqrt{s^2+z^2 }} ~ ds, \\
&=& 2 \int_{0}^{\infty} e^{-\sqrt{s^2+z^2 }} ~ ds, \label{eq:Q2db}\\
&=& 2|z| \int_{0}^{\infty} e^{-|z|\sqrt{1+w ^2 }} ~  dw, \label{eq:Q2dc}\\
&=& 2|z|e^{-|z|}  \int_{0}^{\infty}(1+u)  \frac{ e^{-|z|u}}  {\sqrt{u^2+2u}}~  du, \label{eq:Q2dd} \\
& = & -2|z| \int_0^{\pi/2} \exp{-|z| \sec \theta} \sec \theta~ d\theta, \label{eq:Q2de}
\end{eqnarray}
\end{subequations}
where (\ref{eq:Q2db}) comes from symmetry, (\ref{eq:Q2dc}) comes from letting $s = |z|w$, (\ref{eq:Q2dd}) comes from letting $\sqrt{1+w^2} = u + 1$, and (\ref{eq:Q2de}) comes from the trigonometric substitution $w=\tan \theta$.

From an asymptotic expansion of (\ref{eq:Q2dd}), we find that for small $|z|$,
\begin{equation}
\label{eq:smallz}
Q_{2D} (z)  \approx 2 + \left( \gamma - 1/2 - \ln 2 \right) z^2  +{z}^{2}\ln 
 |z| + {\cal O} \left ( z^4 \ln|z| \right ),
\end{equation}
whereas for large $|z|$,
\begin{equation}
\label{eq:largez}
Q_{2D} (z)  \approx
\sqrt {2 \pi |z|} ~ e^{-|z|} \left [ 1 + \frac{3}{8|z|} - \frac{15}{128z^2} + {\cal O} \left ( 1/|z|^3  \right )
\right ].
\end{equation}
In our numerical study, it is important to have an efficient method of computing values of $Q_{2D}$. In practice, we use (\ref{eq:smallz}) for small $|z|$, (\ref{eq:largez}) for large $|z|$, and for intermediate values of $|z|$ we interpolate from a lookup table pre-computed using~(\ref{eq:Q2de}).

\begin{figure}
\centerline{\resizebox{\textwidth}{!}{\includegraphics{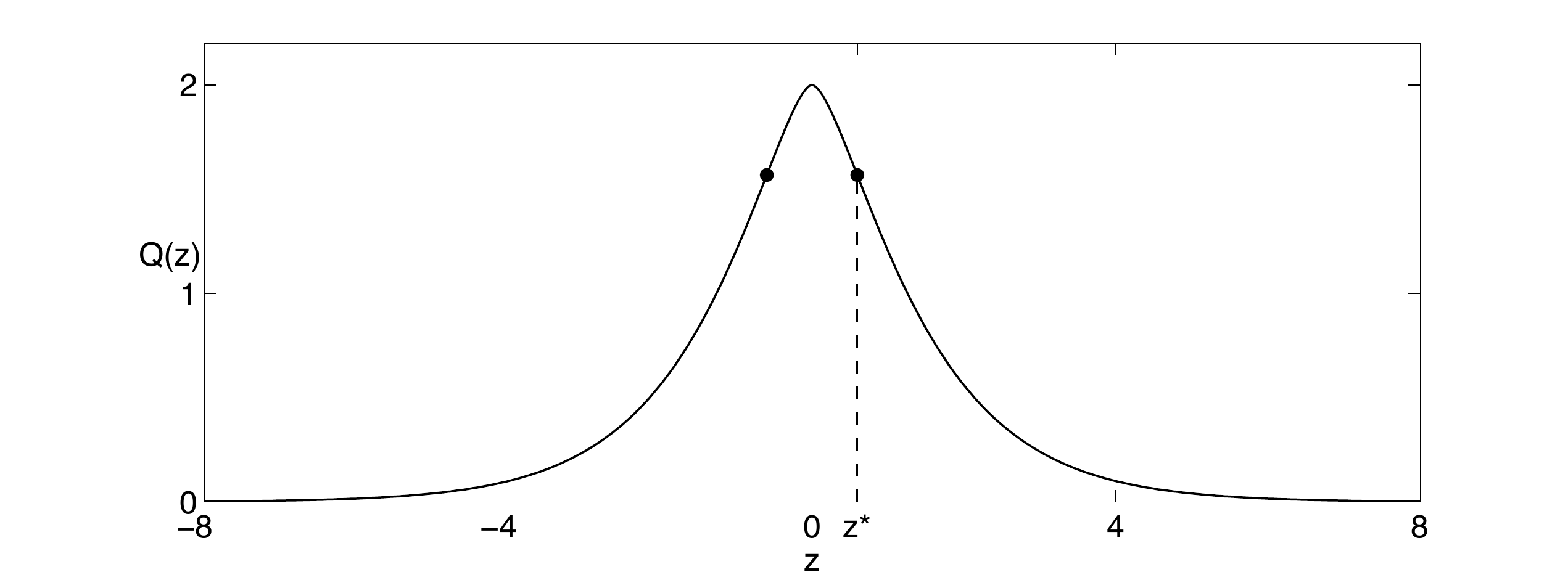}}}
\caption{Quasi-two-dimensional Laplace potential $Q_{2D}$ given by (\ref{eq:q2dlaplace}). $Q_{2D}$ is horizontal at $z=0$, and monotonically decreasing in $|z|$. There are inflection points at $\pm z^*$ given by (\ref{eq:zstar}). \label{fig:Q2d}}
\end{figure}

The potential $Q_{2D}(z)$ is shown in Figure \ref{fig:Q2d}. Note that $Q_{2D}(z)$ is horizontal at $z=0$, and monotonically decreasing in $|z|$. The negative of the slope $\chi = -Q^\prime_{2D}$ reaches a maximum of
\begin{equation}
\label{eq:zstar}
\chi_{max} \approx 0.93305 \mathrm{\ at\ } z_*\approx 0.59505.
\end{equation}
The quantity $\chi_{max}$ plays a key role in our analysis of minimizers below.

The Fourier transform of $Q_{2D}(z)$ can be evaluated exactly using the integral definition (\ref{eq:Q2dc}) and interchanging the order of integration of $z$ and $w$ to obtain
\begin{equation}
\widehat{Q}_{2D}(k) ={\frac {2\pi }{ \left( 1+{k}^{2} \right) ^{3/2}}},
\end{equation}
which we note is positive, so local minimizers are global minimizers per the discussion in Section \ref{sec:minimizers}.

\subsection{Gravitational potential on the half-line with the quasi-2D-Laplace potential}
\label{sec:grav2}

We model a quasi-two-dimensional biological swarm with repulsive social interactions of Laplace type and subject to an exogenous gravitational potential, $F(x) = gx$. The spatial coordinate $x \geq 0$ describes the elevation above ground. Consequently, $\Om$ is the semi-infinite interval $0 \leq x < \infty$.

From Section \ref{sec:grav}, recall that for the one-dimensional model,
\begin{equation}
\brho(x) = M \delta(x),
\end{equation}
is a minimizer for some $M$, corresponding to all swarm members pinned by gravity to the ground. We consider this same solution as a candidate minimizer for the quasi-two-dimensional problem. In this case, $\brho$ above is actually a minimizer for any mass $M$. To see this, we can compute $\Lambda(x)$,
\begin{equation}
\Lambda(x) \equiv \int_\Om Q(x-y)  \brho(y)\,dy +F(x) = MQ_{2D}(x)+gx.
\end{equation}
Since $\Lambda^\prime(0) = g > 0$, $\Lambda$ increases away from the origin and hence $\brho$ is at least a swarm minimizer.

\begin{figure}
\centerline{\resizebox{\textwidth}{!}{\includegraphics{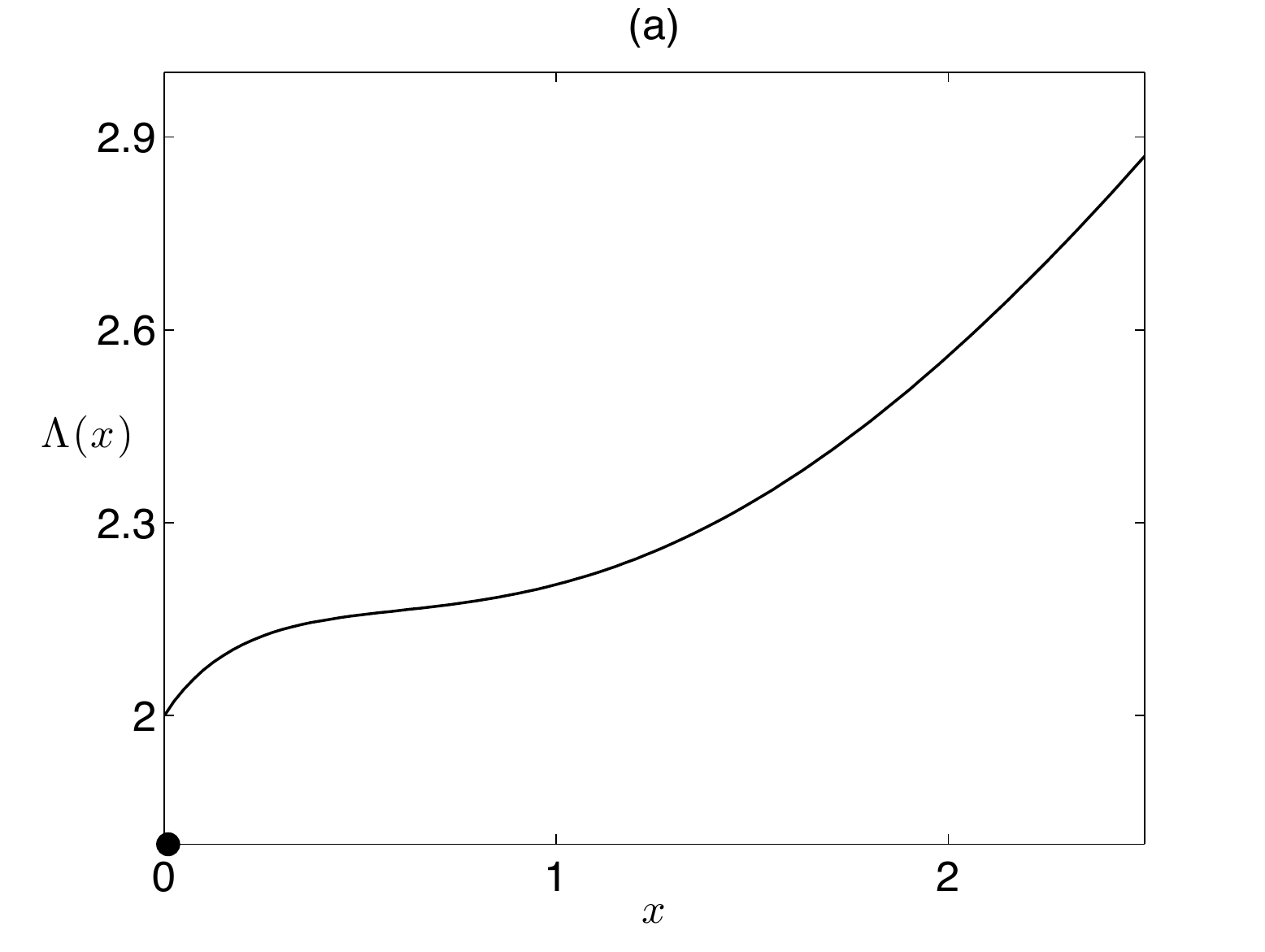} \includegraphics{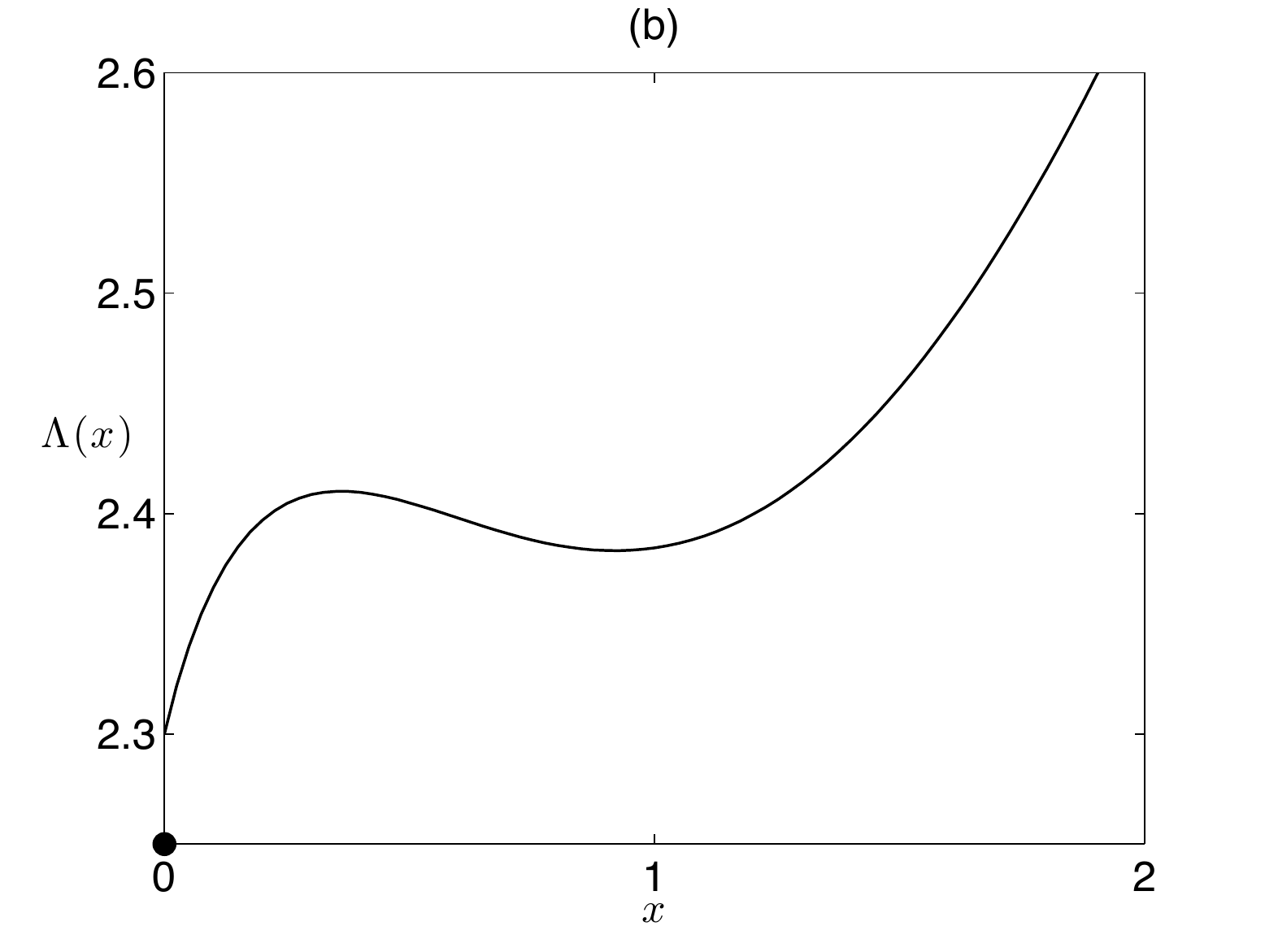}}}
\centerline{\resizebox{\textwidth}{!}{\includegraphics{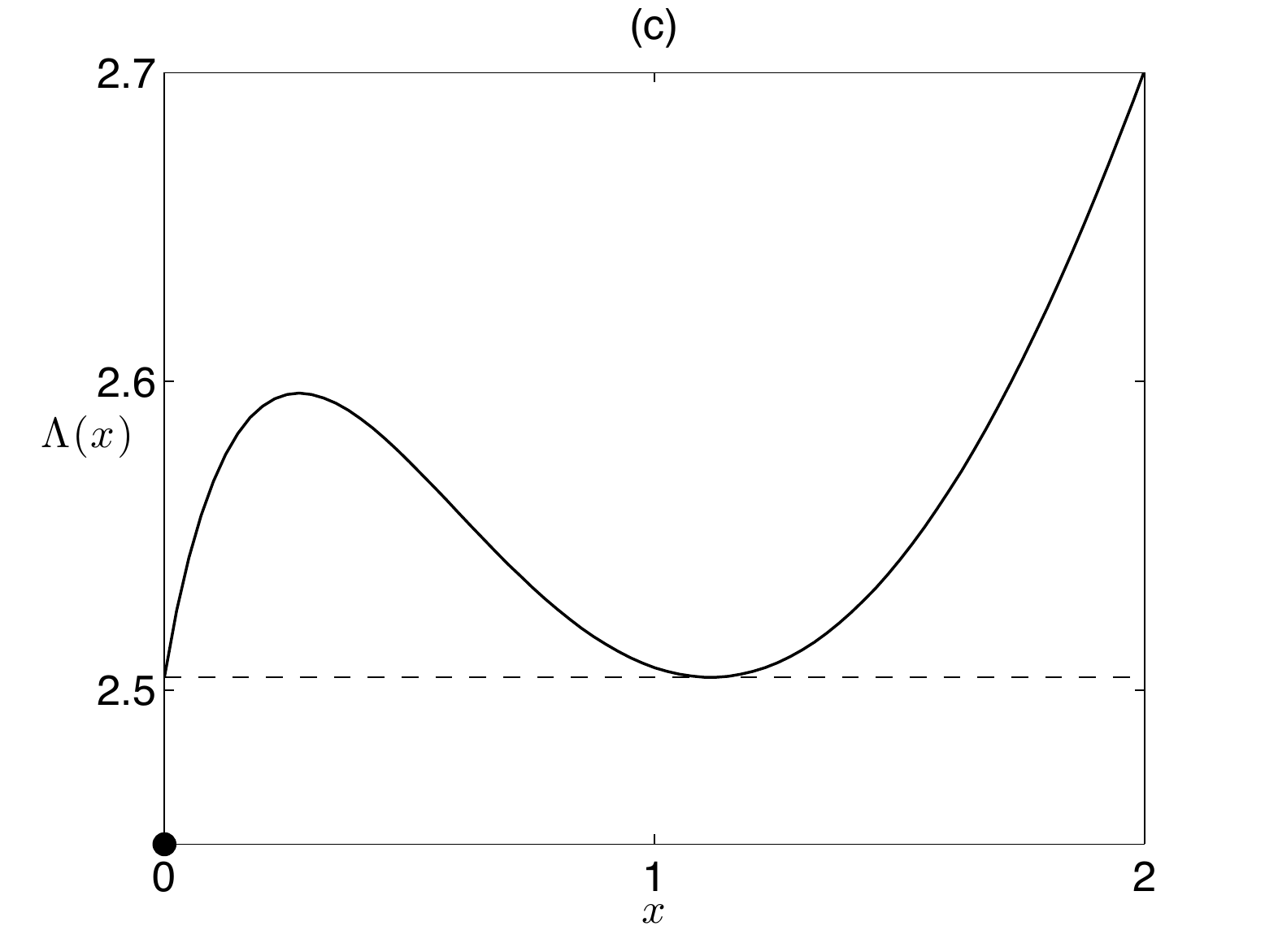} \includegraphics{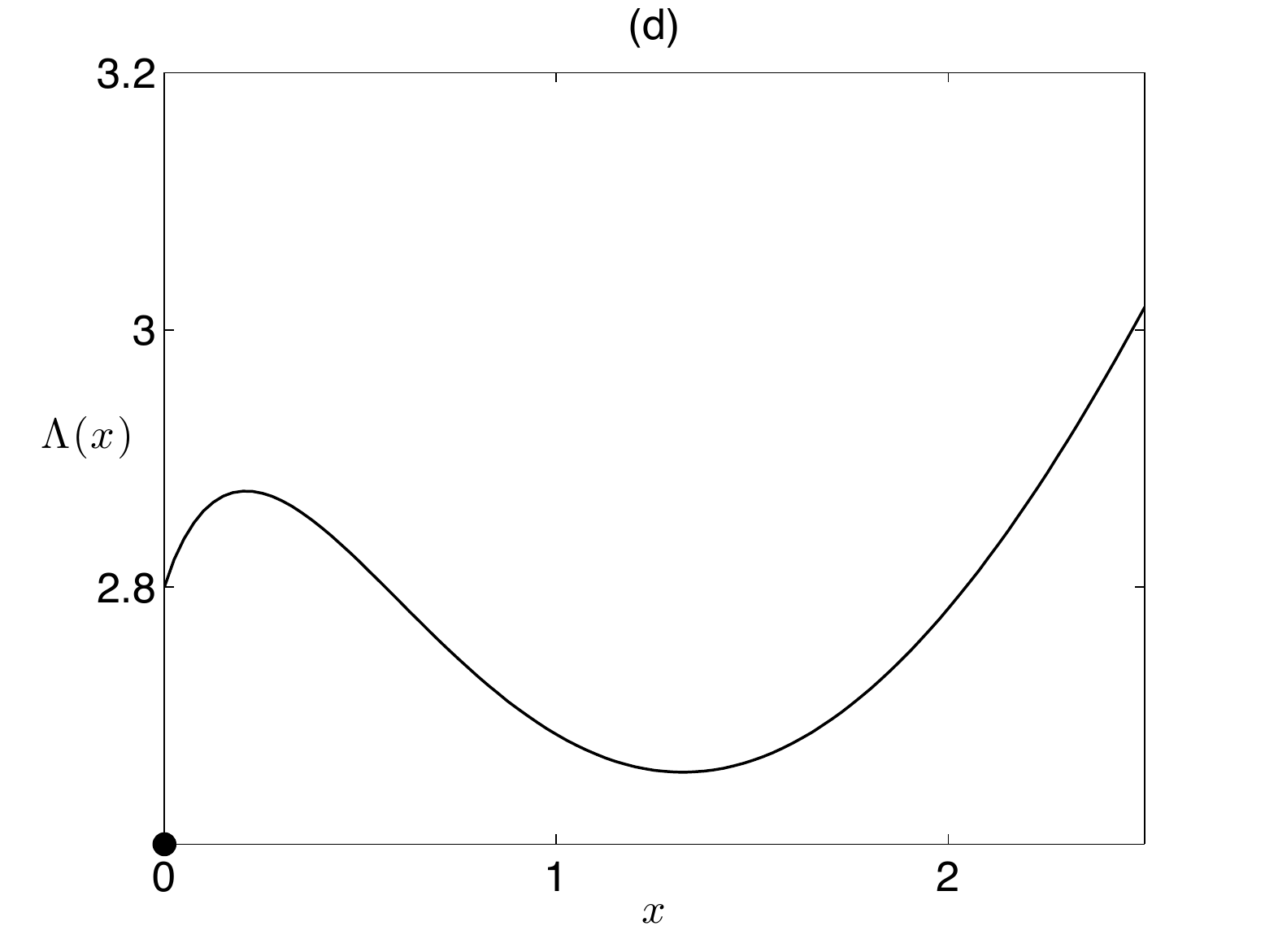}}}
\caption{$\Lambda(x)$ in (\ref{eq:Lambda}) for the half-line $\Omega=[0,\infty]$ with exogenous gravitational potential $F(x) = gx$ with $g=1$,  and endogenous interactions given by the quasi-two-dimensional Laplace potential (\ref{eq:q2dlaplace}). (a) $M = 1 < M_1$, in which case the $\delta$-concentration at the $x=0$ (represented by the dot) is a global minimizer and (we believe) a global attractor. (b) $M_2 > M = 1.15 > M_1$, in which case the $\delta$-concentration is a global minimizer but not a global attractor. Another equilibrium solution can be found by transferring a small amount of mass from the origin to the local minimum in $\Lambda(x)$. (c) $M = M_2 \approx 1.25204$. This is the boundary case; for $M > M_2$, the $\delta$-concentration at the origin is no longer a global minimizer, although it remains a swarm minimizer. (d) $M = 1.4 > M_2$, in which case the $\delta$-concentration at the origin is now only a swarm minimizer.\label{fig:lambda}}
\end{figure}

In fact, if $M < M_1 \equiv g/\chi_{max}$, $\brho(x)$ is a global minimizer because 
\begin{equation}
\Lambda^\prime (x) = MQ_{2D}^\prime(x)+g > g - M \chi_{max} > 0,
\end{equation}
which guarantees that $\Lambda(x)$ is strictly increasing for $x > 0$ as shown in Figure \ref{fig:lambda}(a). Because it is strictly increasing, $\Lambda(x) > \Lambda(0) = \lambda$ for $x>0$. Given this fact, and additionally, since $W_2 > 0$ as previously shown, $\brho$ is a global minimizer. This means that if an infinitesimal amount of mass is added anywhere in the system, it will descend to the origin. Consequently, we believe this solution is the global attractor (though we have not proven this). 

Note that while the condition $ M < M_1$ is sufficient for $\brho$ to be a global minimizer, it is not necessary.
As alluded above, it is not necessary that $\Lambda(x) $ be strictly increasing, only 
that $\Lambda(x) >\Lambda(0)$ for $x>0$. This is the case for for $M<  M_2  \equiv g/\chi_2$, where $\chi_2 \approx 0.79870$. Figure \ref{fig:lambda}(b) shows a case when $ M_1 <M < M_2$. Although $\Lambda(x) >\Lambda(0)$ for $x>0$,  $\Lambda(x)$ has a local minimum. In this situation, although the solution with the mass concentrated at the origin is a global minimizer, it is {\em not} a global attractor. We will see that a small amount of mass added near the local minimum of  $\Lambda(x)$ will create a swarm minimizer, which is dynamically stable to perturbations.

Figure \ref{fig:lambda}(c) shows the critical case when $M =M_2$. In this case the local minimum of $\Lambda(x)$ at  $x=x_2 \approx 1.11436$ satisfies $\Lambda(x_2)=\Lambda(0)$ and $\Lambda^\prime(x_2)=0$.  Figure \ref{fig:lambda}(d) shows the case when $M >M_2$ and now $\Lambda(x)<\Lambda(0)$ in the neighborhood of the minimum. In this case the solution with the mass concentrated at the origin is only a swarm minimizer; the energy of the system can be reduced by transporting some of the mass at the origin to the neighborhood of the  local minimum.

%
%

\begin{figure}
\centerline{\resizebox{!}{2.25in}{\includegraphics{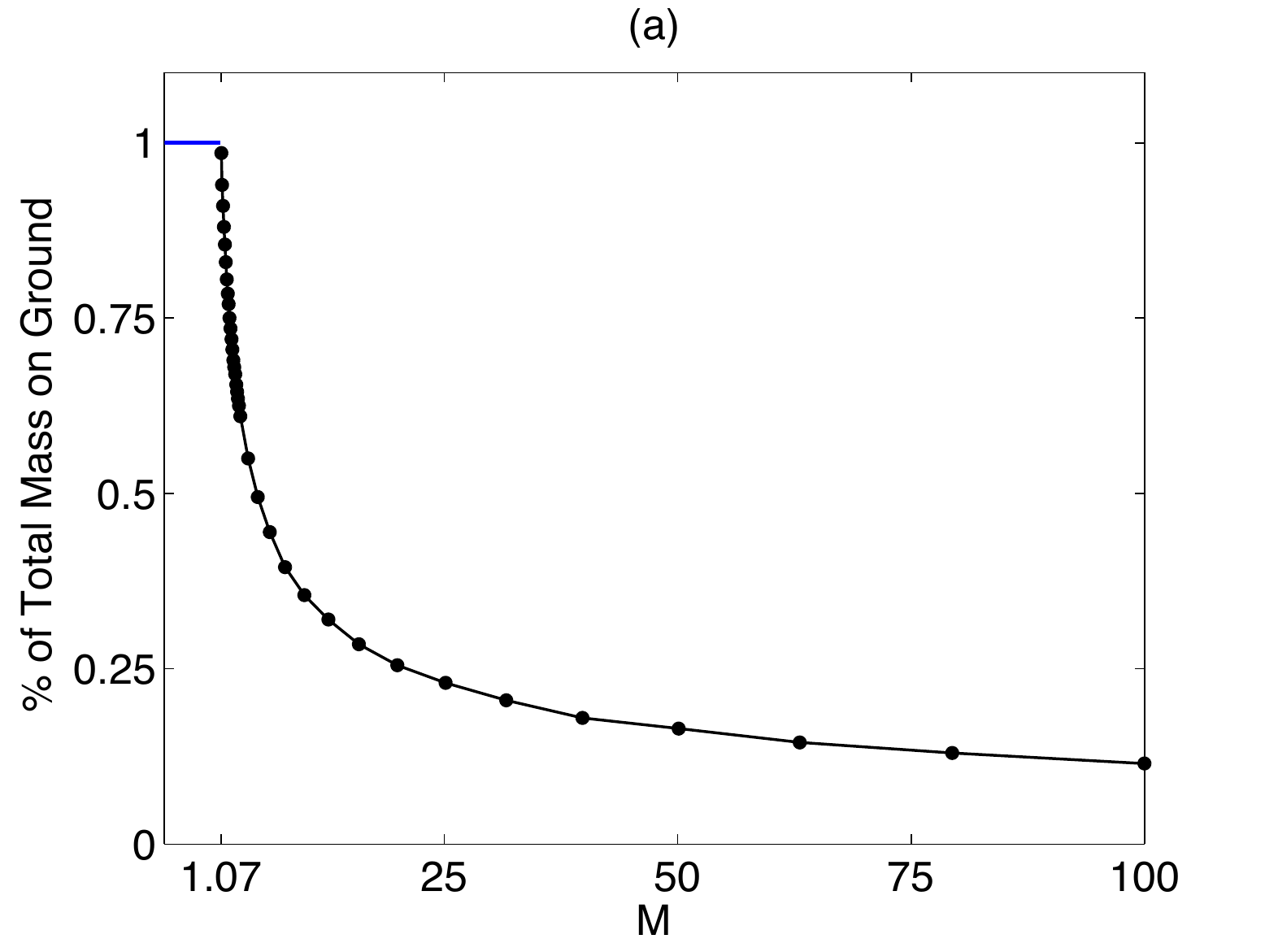}}}
\centerline{\resizebox{!}{2.25in}{\includegraphics{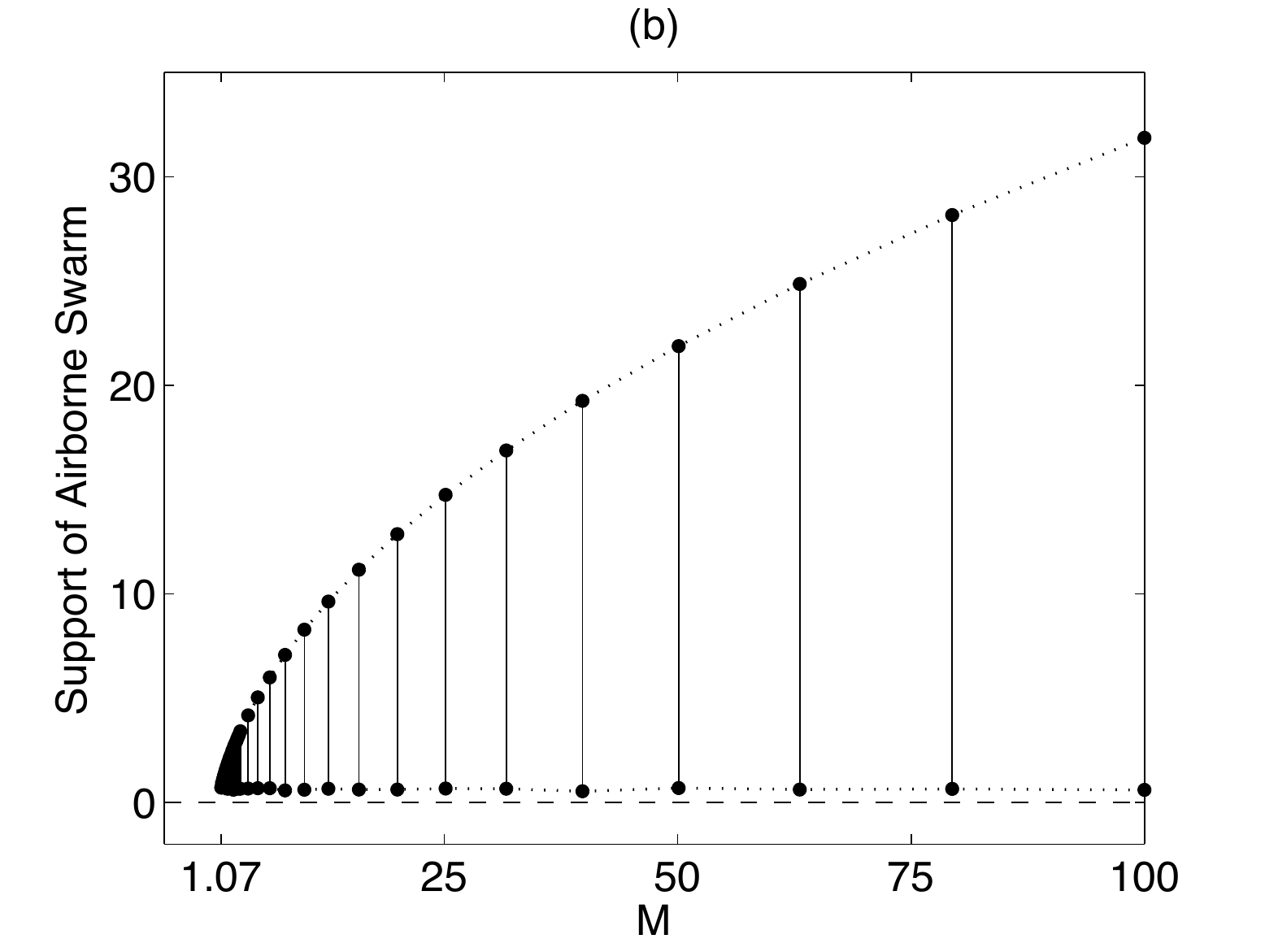}}}
\caption{Numerical simulations of (\ref{eq:discretesystem}) for the half-line $\Omega=[0,\infty]$ with exogenous gravitational potential $F(x) = gx$ and endogenous interactions given by the quasi-two-dimensional Laplace potential (\ref{eq:q2dlaplace}). (a) Mass on the ground as a percentage of the total swarm mass. The horizontal blue line indicates (schematically) that for $M < M_1$, the equilibrium consists of all mass concentrated at the origin. As mass is increased through $M_1$, a classical swarm exists in the air separated from the origin, and the proportion of mass located on the ground decreases monotonically. (b) Support of the airborne swarm, which exists only for $M > M_1$; the lower and upper data represent the coordinates of the bottom and top of the swarm, respectively. The bottom of the swarm is located a finite distance from the origin, which is represented as the horizontal dotted line. As mass is increased, the span of the swarm increases monotonically, with the location of the bottom remaining constant within the error of our Lagrangian numerical approximation. For all computations, we take $g=1$ and $N = 200$ swarm members in the discrete simulation. We use a linear spacing in $M$ for larger values of $M$, but a logarithmic spacing for values close to $M_1$ in order to resolve the bifurcation.\label{fig:quasi2dnumerics}}
\end{figure}

When $ M >M_1$ it is possible to construct a continuum of swarm minimizers. We have conducted a range of simulations for varying $M$ and have measured two basic properties of the solutions. We set $g = 1$ and use $N = 200$ in all simulations of the discrete system. Initially, all the swarm members are high above the ground and we evolve the simulation to equilibrium. Figure \ref{fig:quasi2dnumerics}(a) measures the mass on the ground as a percentage of the total swarm mass. The horizontal blue line indicates (schematically) that for $M < M_1$, the equilibrium consists of all mass concentrated at the origin; as discussed above, this state is the global minimizer and (we believe) the global attractor. As mass is increased through $M_1$, the equilibrium is a swarm minimizer consisting of a classical swarm in the air separated from the origin, and some mass concentrated on the ground. As $M$ increases, the proportion of mass located on the ground decreases monotonically. Figure \ref{fig:quasi2dnumerics}(b) visualizes the support of the airborne swarm, which exists only for $M > M_1$; the lower and upper data represent the coordinates of the bottom and top of the swarm, respectively. As mass is increased, the span of the swarm increases monotonically.

\begin{figure}
\centerline{\resizebox{\textwidth}{!}{\includegraphics{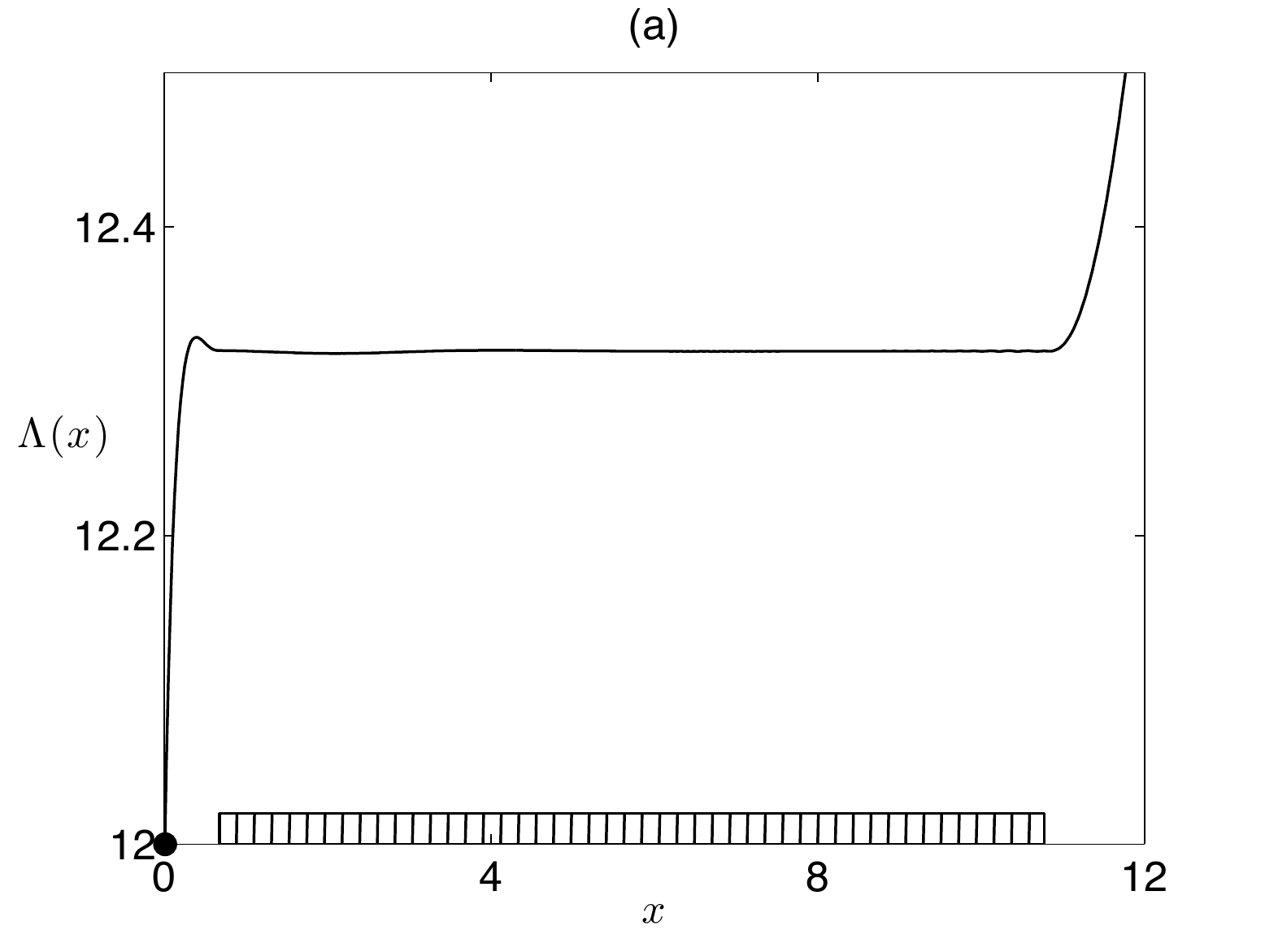} \includegraphics{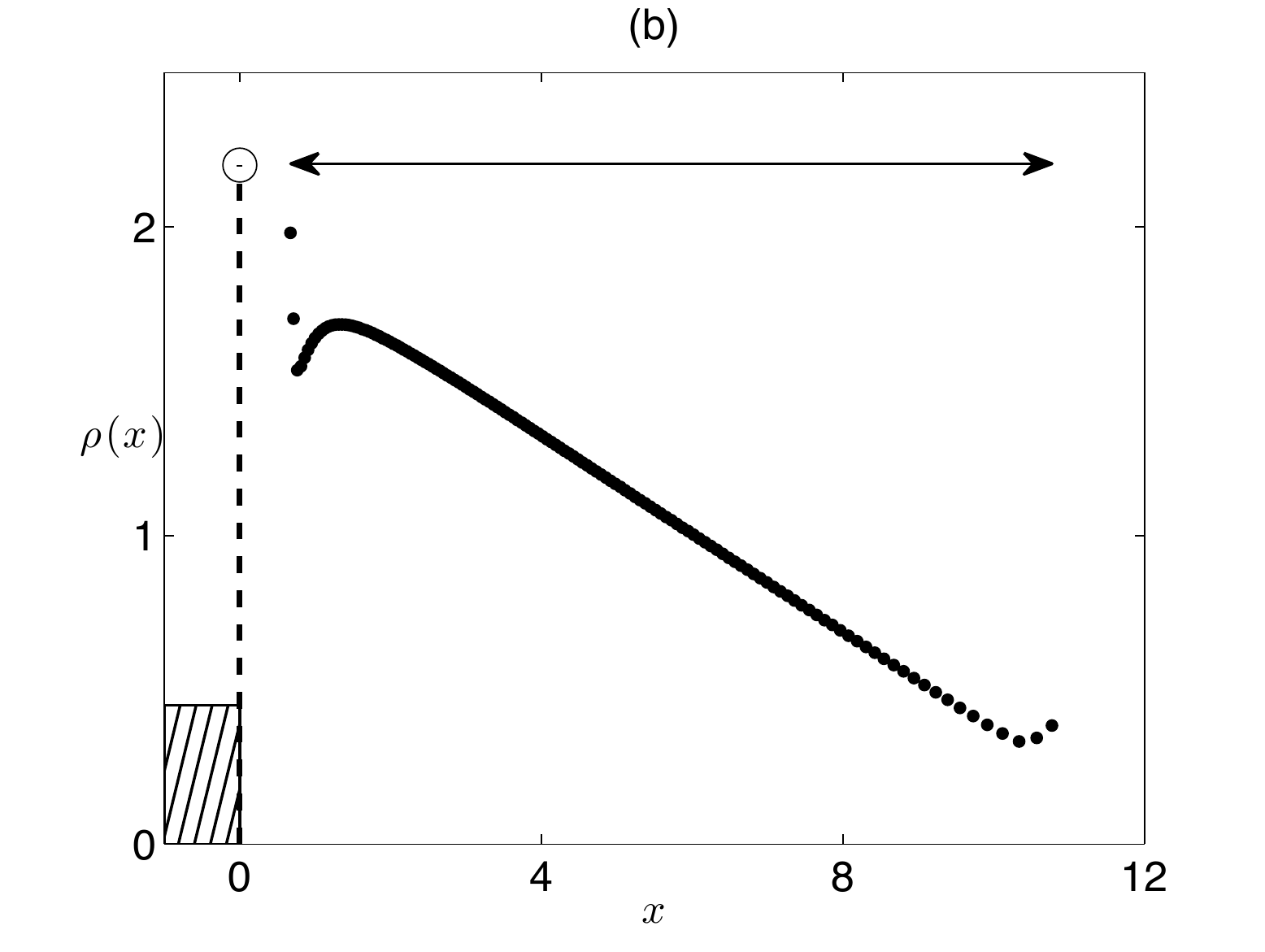}}}
\centerline{\resizebox{\textwidth}{!}{\includegraphics{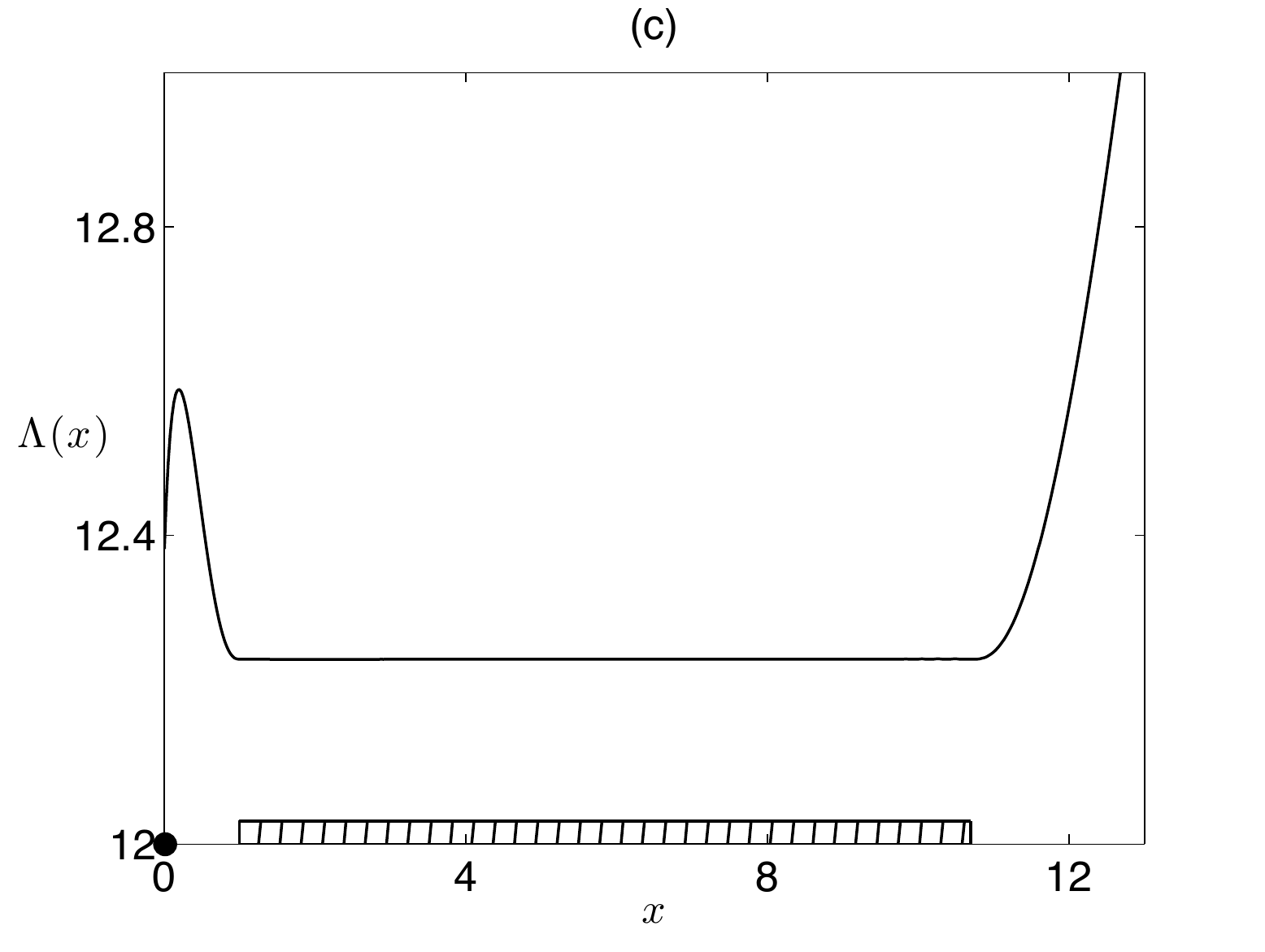} \includegraphics{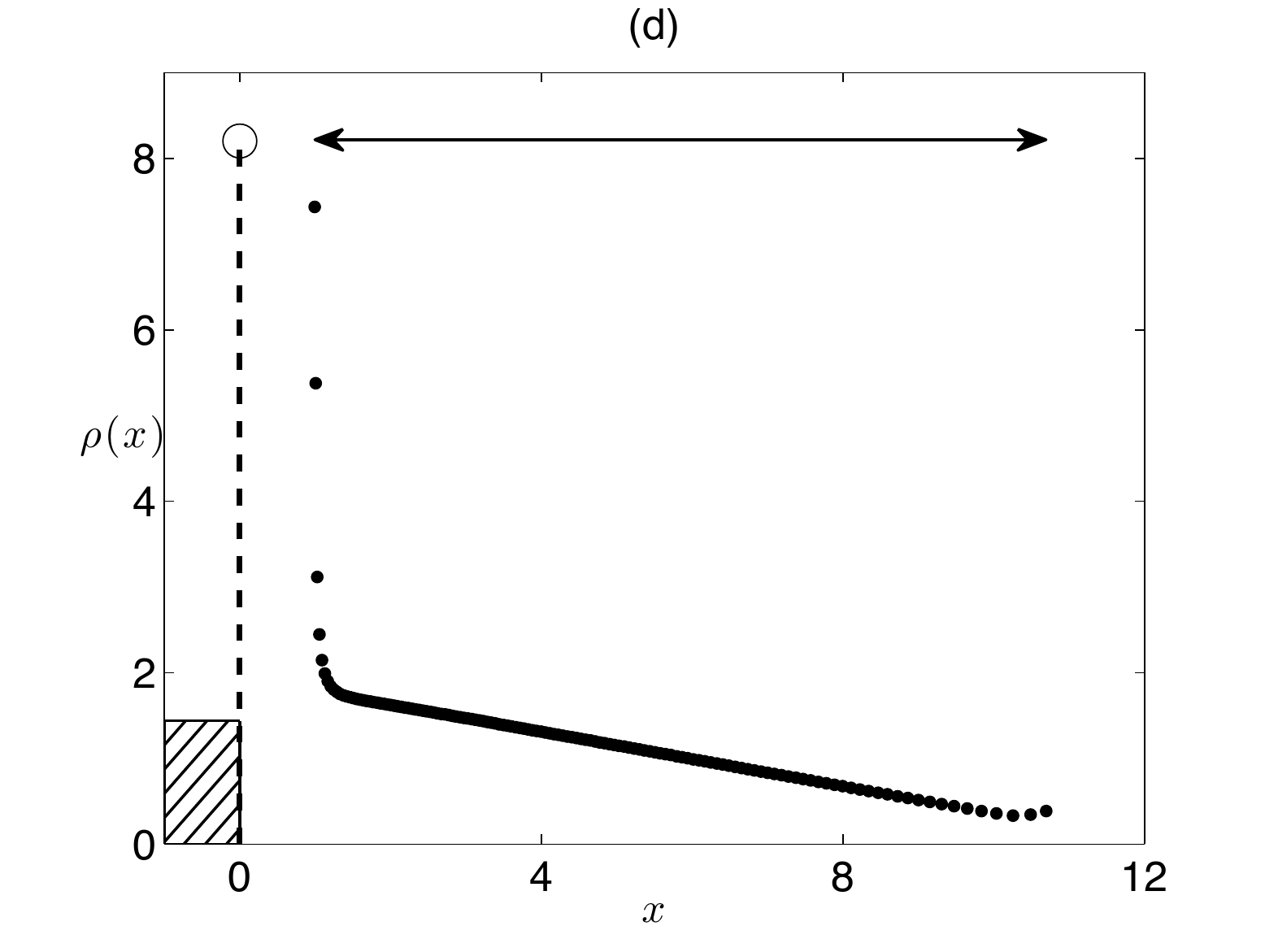}}}
\caption{Two example swarm minimizers for the quasi-two-dimensional Laplace potential (\ref{eq:q2dlaplace}) on the half-line $\Omega=[0,\infty]$ with exogenous gravitational potential $F(x) = gx$. We take $g=1$ and $N=200$ swarm members for our simulations, but choose different initial conditions. Each minimizer consists of two swarm components, namely a concentration of mass at the origin and a finite-sized swarm separated from the origin by a gap. (ab) $29.5\%$ of the mass is in the grounded component. (a) Shows $\Lambda(x)$. The dot and cross-hatched bar schematically indicate the support of the two components. Since $\lambda_0 \equiv \Lambda(0) < \lambda_1$ where $\lambda_1$ is $\Lambda(x)$ in the airborne component, we conclude that the total energy could be reduced by transporting mass from the air to the ground. (b) The swarm density $\brho(x)$. The ``lollipop'' represents a superposition of $59$ swarm members, having total mass $59 \cdot 1/200$. Density within the airborne component is given only at the Lagrangian gridpoints. The horizontal arrow helps demarcate its support. (cd) Like (ab), but here $32.5\%$ of the mass is contained in the grounded group. Since $\lambda_0 < \lambda_1$, the total energy could be reduced by transporting mass from the ground to the air. Note that although both states could have their energy reduced by transporting mass between components, each one is still a dynamically stable equilibrium. We believe that these are just two representative examples of a continuum of dynamically stable swarm equilibria all with the same total mass.\label{fig:lambda2}}
\end{figure}

As established above, when $M > M_1$, swarm minimizers exist with two components. In fact, there is a continuum of swarm minimizers with different proportions of mass in the air and on the ground. Which minimizer is obtained in simulation depends on initial conditions. Figure \ref{fig:lambda2} shows two such minimizers for $g = 1$ and $M= 15 > M_2$,  and the associated values of $\Lambda(x)$ (each obtained from a different initial condition). Recalling that for a swarm minimizer, each connected component of the swarm, $\lambda$ is constant, we define $\Lambda(x) = \Lambda(0) = \lambda_0$ for the grounded component and $\Lambda(x) = \lambda_1$ for the airborne component. In Figure \ref{fig:lambda2}(ab), $29.5\%$ of the mass is contained in the grounded component. In this case, $\lambda_0 < \lambda_1$ indicating that the total energy could be reduced by transporting swarm members from the air to the ground. In contrast, in Figure \ref{fig:lambda2}(cd), $32.5\%$ of the mass is contained in the grounded component. In this case, $\lambda_1 < \lambda_0$ indicating that the total energy could be reduced by transporting swarm members from the ground to the air. Note that by continuity, we believe a state exists where $\lambda_0 = \lambda_1$, which would correspond to a global minimizer. However, this state is clearly not a global attractor and hence will not necessarily be achieved in simulation.

We've demonstrated that for $M>M_1$ one can construct a continuum of swarm minimizers with a gap between grounded and airborne components, and that for $M>M_2$ these solutions can have a lower energy than the state with the density concentrated solely on the ground. By contrast with the one-dimensional system of Section \ref{sec:grav} in which no gap is observed, these gap states appear to be the generic configuration for sufficiently large mass in the quasi-two-dimensional system. We conclude that dimensionality is crucial element for the formation of the bubble-like shape of real locust swarms.

\section{Conclusions}
\label{sec:conc}

In this paper we deeveloped a framework for studying equilibrium solutions for swarming problems. We related the discrete swarming problem to an associated continuum model. This continuum model has an energy formulation which enables analysis equilibrium solutions and their stability. We derived conditions for an equilibrium solution to be a local minimizer, a global minimizer, and/or a swarm minimizer, that is, stable to infinitesimal Lagrangian deformations of the mass.

We found many examples of compactly supported equilibrium solutions, which may be discontinuous at the boundary of the support. In addition, when a boundary of the support coincides with the domain boundary, a minimizer may contain a $\delta$-concentration there. For the case of exogenous repulsion modeled by the Laplace potential, we computed three example equilibria. On a bounded domain, the minimizer is a constant density profile with $\delta$-functions at each end. On a half-line with an exogenous  gravitational potential, the minimizer is a compactly supported linear density profile with a $\delta$-function at the origin. In free space with an exogenous quadratic potential, the minimizer is a compactly supported inverted parabola with jump discontinuities at the endpoints. Each of the aforementioned solutions is also a global minimizer.

To extend the results above, we also found analytical solutions for exogenous attractive-repulsive forces, modeled with the Morse potential. In the case that the social force was in the catastrophic statistical mechanical regime, we found a compactly supported solution whose support is independent of the total population mass. This means that within the modeling assumptions, swarms become denser with increasing mass. For the case of an H-stable social force, there is no equilibrium solution on an infinite domain. On a finite domain, mass is partitioned between a classical solution in the interior and $\delta$-concentrations on the boundary.

We recall that for the locust model of \cite{TopBerLog2008} (see Figure \ref{fig:locust}) a concentration of locusts occurs on the ground, with a seemingly classical component above, separated by a gap. None of the one-dimensional solutions (for the Laplace and Morse potentials) discussed above contain a gap, that is, multiple swarm components that are spatially disconnected, suggesting that this configuration is intrinsically two-dimensional. To study this configuration, we computed a quasi-two-dimensional potential corresponding to a horizontally uniform swarm. We demonstrated numerically that for a wide range of parameters, there exists a continuous family of swarm minimizers that consist of a concentration on the ground and a disconnected, classical component in the air, reminiscent of our earlier numerical studies of a discrete locust swarm model.

We believe that the analytical solutions we found provide a sampling of the rich tapestry of equilibrium solutions that manifest in the general model we have considered, and in nature. We hope that these solutions will inspire further analysis and guide future modeling efforts.

\section*{Acknowledgments}

CMT acknowledges support from the NSF through grants DMS-0740484 and DMS-1009633. AJB gratefully acknowledges the support from the NSF through grants DMS-0807347 and DMS-0730630, and the hospitality of Robert Kohn and  the Courant Institute of Mathematical Sciences. We both wish to thank the Institute for Mathematics and Its Applications where portions of this work were completed.


\begin{thebibliography}{10}

\bibitem{Alb1967}
{\sc F.~O. Albrecht}, {\em Polymorphisme Phasaire et Biologie des Acridiens
  Migrateurs}, Les Grands Probl{\`{e}}mes de la Biologie, Masson, Paris, 1967.

\bibitem{BenCohLev2000}
{\sc E.~Ben-{J}acob, I.~Cohen, and H.~Levine}, {\em Cooperative
  self-organization of microorganisms}, Adv. Phys., 49 (2000), pp.~395--554.

\bibitem{BerBra2010}
{\sc A.~L. Bertozzi and J.~Brandman}, {\em Finite-time blow-up of
  ${L}^\infty$-weak solutions of an aggregation equation}, Comm. Math. Sci., 8
  (2010), pp.~45--65.

\bibitem{BerCarLau2009}
{\sc A.~L. Bertozzi, J.~A. Carrillo, and T.~Laurent}, {\em Blow-up in
  multidimensional aggregation equations with mildly singular interaction
  kernels}, Nonlinearity, 22 (2009), pp.~683--710.

\bibitem{BerLau2007}
{\sc A.~L. Bertozzi and T.~Laurent}, {\em Finite-time blow-up of solutions of
  an aggregation equation in $\mathbb{R}^n$}, Comm. Math. Phys., 274 (2007),
  pp.~717--735.

\bibitem{BerLau2009}
\leavevmode\vrule height 2pt depth -1.6pt width 23pt, {\em The behavior of
  solutions of multidimensional aggregation equations with mildly singular
  interaction kernels}, Chinese Ann. Math. Ser. B, 30 (2009), pp.~463--482.

\bibitem{BerLauRos2010}
{\sc A.~L. Bertozzi, T.~Laurent, and J.~Rosado}, {\em ${L}^p$ theory for the
  multidimensional aggregation equation}, Comm. Pure App. Math.,  (2010).

\bibitem{BodVel2005}
{\sc M.~Bodnar and J.~J. Velasquez}, {\em Derivation of macroscopic equations
  for individual cell-based models: A formal approach}, Math. Meth. Appl. Sci.,
  28 (2005), pp.~1757--1779.

\bibitem{BodVel2006}
\leavevmode\vrule height 2pt depth -1.6pt width 23pt, {\em An
  integro-differential equation arising as a limit of individual cell-based
  models}, J. Diff. Eq., 222 (2006), pp.~341--380.

\bibitem{Bre1954}
{\sc C.~M. Breder}, {\em Equations descriptive of fish schools and other animal
  aggregations}, Ecol., 35 (1954), pp.~361--370.

\bibitem{CarDifFig2010}
{\sc J.~A. Carrillo, M.~DiFrancesco, A.~Figalli, T.~Laurent, and D.~Slepcev},
  {\em Global-in-time weak measure solutions, finite-time aggregation, and
  confinement for nonlocal interaction equations}, Duke Math. J.,  (2010).

\bibitem{ChuDOrMar2007}
{\sc Y.~L. Chuang, M.~R. D'Orsogna, D.~Marthaler, A.~L. Bertozzi, and L.~S.
  Chayes}, {\em State transitions and the continuum limit for a 2{D}
  interacting, self-propelled particle system}, Physica D, 232 (2007),
  pp.~33--47.

\bibitem{CouKraJam2002}
{\sc I.~D. Couzin, J.~Krause, R.~James, G.~D. Ruxton, and N.~R. Franks}, {\em
  Collective memory and spatial sorting in animal groups}, J. Theor. Biol., 218
  (2002), pp.~1--11.

\bibitem{DOrChuBer2006}
{\sc M.~R. D'Orsogna, Y.~L. Chuang, A.~L. Bertozzi, and L.~Chayes}, {\em
  Self-propelled particles with soft-core interactions: Patterns, stability,
  and collapse}, Phys. Rev. Lett., 96 (2006), pp.~104302.1--104302.4.

\bibitem{EdeWatGru1998}
{\sc L.~Edelstein-Keshet, J.~Watmough, and D.~Gr{\"{u}}nbaum}, {\em Do
  travelling band solutions describe cohesive swarms? {A}n investigation for
  migratory locusts}, J. Math. Bio., 36 (1998), pp.~515--549.

\bibitem{FelRao2010}
{\sc K.~Fellner and G.~Raoul}, {\em Stability of stationary states of non-local
  equations with singular interaction potentials}, Math. Comp. Model.,  (2010).

\bibitem{FelRao2010b}
\leavevmode\vrule height 2pt depth -1.6pt width 23pt, {\em Stable stationary
  states of non-local interaction equations}, Math. Mod. Meth. Appl. Sci.,
  (2010).

\bibitem{GreCha2004}
{\sc G.~Gr{\'{e}}goire and H.~Chat{\'{e}}}, {\em Onset of collective and
  cohesive motion}, Phys. Rev. Lett., 92 (2004), pp.~025702.1--025702.4.

\bibitem{GreChaTu2001}
{\sc G.~Gr{\'{e}}goire, H.~Chat{\'{e}}, and Y.~Tu}, {\em Active and passive
  particles: Modeling beads in a bacterial bath}, Phys. Rev. E, 64 (2001),
  pp.~011902.1--011902.7.

\bibitem{GreChaTu2003}
\leavevmode\vrule height 2pt depth -1.6pt width 23pt, {\em Moving and staying
  together without a leader}, Physica D, 181 (2003), pp.~157--170.

\bibitem{HolPut2005}
{\sc D.~D. Holm and V.~Putkaradze}, {\em Aggregation of finite size particles
  with variable mobility}, Phys. Rev. Lett., 95 (2005), pp.~226106.1 --
  226106.4.

\bibitem{HolPut2006}
\leavevmode\vrule height 2pt depth -1.6pt width 23pt, {\em Formation of clumps
  and patches in self-aggregation of finite-size particles}, Physica D, 220
  (2006), pp.~183--196.

\bibitem{LevTopBer2009}
{\sc A.~J. Leverentz, C.~M. Topaz, and A.~J. Bernoff}, {\em Asymptotic dynamics
  of attractive-repulsive swarms}, SIAM J. Appl. Dyn. Sys., 8 (2009),
  pp.~880--908.

\bibitem{LevRapCoh2001}
{\sc H.~Levine, W.~J. Rappel, and I.~Cohen}, {\em Self-organization in systems
  of self-propelled particles}, Phys. Rev. E, 63 (2001),
  pp.~017101.1--017101.4.

\bibitem{LukLiEde2010}
{\sc R.~Lukeman, Y.~X. Li, and L.~Edelstein-Keshet}, {\em Inferring individual
  rules from collective behavior}, Proc. Natl. Acad. Sci., 107 (2010),
  pp.~12576--12580.

\bibitem{MilYan2008}
{\sc P.~A. Milewski and X.~Yang}, {\em A simple model for biological
  aggregation with asymmetric sensing}, Comm. Math. Sci., 6 (2008),
  pp.~397--416.

\bibitem{MogEde1999}
{\sc A.~Mogilner and L.~Edelstein-Keshet}, {\em A non-local model for a swarm},
  J. Math. Bio., 38 (1999), pp.~534--570.

\bibitem{MogEdeBen2003}
{\sc A.~Mogilner, L.~Edelstein-Keshet, L.~Bent, and A.~Spiros}, {\em Mutual
  interactions, potentials, and individual distance in a social aggregation},
  J. Math. Bio., 47 (2003), pp.~353--389.

\bibitem{OkuLev2001}
{\sc A.~Okubo and S.~A. Levin}, eds., {\em Diffusion and Ecological Problems},
  vol.~14 of Interdisciplinary Applied Mathematics: Mathematical Biology,
  Springer, New York, second~ed., 2001.

\bibitem{ParHam1997}
{\sc J.~K. Parrish and W.~M. Hamner}, eds., {\em Animal Groups in Three
  Dimensions}, Cambridge University Press, Cambridge, UK, 1997.

\bibitem{Rai1989}
{\sc R.~C. Rainey}, {\em Migration and Meteorology: Flight Behavior and the
  Atmospheric Environment of Locusts and other Migrant Pests}, Oxford Science
  Publications, Clarendon Press, Oxford, 1989.

\bibitem{Rao2010}
{\sc G.~Raoul}, {\em Non-local interaction equations: Stationary states and
  stability analysis}.
\newblock Preprint, 2010.

\bibitem{Rue1969}
{\sc D.~Ruelle}, {\em Statistical Mechanics: Rigorous Results}, Mathematical
  Physics Monograph Series, W. A. Benjamin, New York, 1969.

\bibitem{TopBerLog2008}
{\sc C.~M. Topaz, A.~J. Bernoff, S.~Logan, and W.~Toolson}, {\em A model for
  rolling swarms of locusts}, Euro. Phys. J. ST, 157 (2008), pp.~93--109.

\bibitem{TopBer2004}
{\sc C.~M. Topaz and A.~L. Bertozzi}, {\em Swarming patterns in a
  two-dimensional kinematic model for biological groups}, SIAM J. Appl. Math.,
  65 (2004), pp.~152--174.

\bibitem{TopBerLew2006}
{\sc C.~M. Topaz, A.~L. Bertozzi, and M.~A. Lewis}, {\em A nonlocal continuum
  model for biological aggregation}, Bull. Math. Bio., 68 (2006),
  pp.~1601--1623.

\bibitem{Uva1977}
{\sc B.~Uvarov}, {\em Grasshoppers and Locusts}, vol.~2, Cambridge University
  Press, London, UK, 1977.

\bibitem{VicCziBen1995}
{\sc T.~Vicsek, A.~Czirok, E.~Ben~Jacob, I.~Cohen, and O.~Shochet}, {\em Novel
  type of phase-transition in a system of self-driven particles}, Phys. Rev.
  Lett., 75 (1995), pp.~1226--1229.

\bibitem{Vil2003}
{\sc C.~Villani}, {\em Optimal transportation, dissipative {PDE}'s and
  functional inequalities}, in Optimal transportation and applications
  ({M}artina {F}ranca, 2001), vol.~1813 of Lecture Notes in Math., Springer,
  Berlin, 2003, pp.~53--89.

\end{thebibliography}
\end{document}